\documentclass[aps,prd,twocolumn,notitlepage,nofootinbib,,nobibnotes,superscriptaddress,amsmath,amssymb,preprintnumbers]{revtex4-1}

\usepackage{graphicx} 
\usepackage{dcolumn}
\usepackage{bm}        
\usepackage{amssymb}   
\usepackage{adjustbox}  
\usepackage{amsmath,array}
\usepackage{comment}
\usepackage{natbib}
\usepackage{MnSymbol}
\usepackage[hidelinks]{hyperref}
\usepackage{graphicx}
\usepackage{subfig}
\usepackage{color}
\usepackage     [utf8]                  {inputenc}
\usepackage     [T1]                    {fontenc}
\usepackage     [english]               {babel}

\usepackage{epigraph}
\usepackage{etoolbox}
\usepackage{ dsfont }
\usepackage{physics}
\usepackage{siunitx}
\makeatletter
\patchcmd{\epigraph}{\@epitext{#1}}{\itshape\@epitext{#1}}{}{}  
\newcommand*\eqsize{%
\@setfontsize\mysize{9.0}{9.0}%
    }
    \usepackage{multirow}
\makeatother

\usepackage{xfrac}
\usepackage{amsmath,amssymb}
\usepackage{esdiff} 
\usepackage{commath} 
\usepackage{bbm} 
\usepackage{braket}
\usepackage{slashed}

\newcommand{\rmd}{\mathrm{d}}

\newcommand{\aem}{\alpha}
\newcommand{\as}{\alpha_{S}}

\newcommand{\xt}{x_\perp}			
\newcommand*\PTS[1]{{p_{#1}}_\perp}				

\newcommand{\GAMMA}{\frac{40}{9\pi^2} \frac{\alpha_e\,\kappa_g}{4\pi^2}}

\newcommand{\WR}{W_r[p_\perp,Q_s]}

\begin{document}

\date{\today}

\title{Non-equilibrium photons from the \textit{bottom-up} thermalization scenario}

\author{Oscar Garcia-Montero  \\\vspace{2pt} \small \textit{Institut f\"{u}r Theoretische Physik,
                     Universit\"{a}t Heidelberg,}\\ \textit{Philosophenweg 16,
                     69120 Heidelberg, Germany}\normalsize\vspace{2pt}}

\email{garcia@thphys.uni-heidelberg.de}

\begin{abstract} 
In this work, I calculate the $p_\perp$ resolved spectra for the three stages of the \textit{bottom-up} scenario, which are comparable to the thermal contribution, particularly at higher values of the saturation scale $Q_S^2$. Analytical solutions are obtained by including a parametrization of scaling solutions from far-from-equilibrium classical statistical lattice simulations into a small angle kinetic rate. Furthermore, a theoretically motivated ansatz is used to account for near-collinear enhancement of the low-$p_\perp$ radiation. The system is phenomenologically constrained using the charge hadron multiplicities from LHC and RHIC as in previous parametric estimates and fair agreement with the data available for photons was found. I find that for this realistic set of parameters, the contribution from the thermalizing glasma dominates the excess photons.
\end{abstract}

\maketitle
\section{Introduction}

Direct photons are radiated throughout the evolution of a heavy ion collision (HIC) and, due to a lack of final-state interactions, can escape the medium virtually unscathed.  As a consequence, these probes are sensitive to the different stages of the rapidly expanding fireball. In small systems, such as $p+p$ or $p+A$ collisions, direct photons produced are mostly prompt photons, whose invariant yield can be calculated perturbatively \cite{Baranov:2007np, Motyka:2016lta,Gordon:1993qc} or using hybrid approaches \cite{Benic:2018hvb, Benic:2016uku} to account for nuclear modification factors. However, in collisions of large nuclear systems, direct photons in the small transverse momentum range exhibit exponential enhancement, commonly explained by  hydrodynamical models. 
 
  In addition to those findings, the transverse plane anisotropy of the photon multiplicities has been studied~\cite{Chatterjee:2005de}, finding non-vanishing flow coefficients~\cite{Adamczyk:2014yew,Acharya:2018bdy}. This anisotropy is thought to arise from the space-time evolution of the underlying medium. To compute such quantities, hydrodynamical quark-gluon-plasma (QGP) and transport models have been compared to the available data.  Unfortunately, the simultaneous reproduction of the yields and the photon flow coefficients, $v_n$,  has been out of reach for theoretical models~\cite{Chatterjee:2013naa,Shen:2014lpa,vanHees:2014ida}. This challenge has been named the \textit{direct photon puzzle}~\cite{Shen:2015nto}. 
  
  Nevertheless, in those calculations the pre-equilibrium physics of the medium is not accounted for, which leads to the introduction of several uncertainties including the initial conditions for the hydrodynamical evolution. Pre-equilibrium sourced photons are also omitted, and while 
 it is the traditional idea that, because of the small early-times space-time volumes, a pre-equilibrium contribution to the direct photon spectrum is negligible, new results seem to suggest that such a source may contribute on the same order of magnitude as the thermal stages \cite{McLerran:2014hza,McLerran:2014oea,Khachatryan:2018ori}. In a novel estimate, \cite{Berges:2017eom}, thermal and Glasma total photon yields were computed parametrically using the \textit{bottom-up} thermalization scenario by Baier, Mueller, Schiff and Son (BMSS) \cite{Baier:2000sb}. They were found to be comparable to thermal total yields. For this, a  phenomenological matching was performed to account for the energy scale $Q_S$ in the system as well as for the thermalization time, $\tau_{th}$ and temperature, $T_{th}$.

While a full phenomenological simulation that links the initial stage of the collision with the onset of hydrodynamics is still out of reach, broad progress has been achieved to understand its dynamics. The initial stage of the collision is classical and highly non-linear in the gluon fields and after a parametrically short time, the evolution leads to instabilities which overpopulate the gluon fields \cite{Gelis:2010nm,Epelbaum:2013waa,Berges:2014yta}. 
Using classical statistical simulations, it was shown that an over-occupied Glasma approaches a non-thermal fixed point \cite{Berges:2013eia,Orioli:2015dxa}, and by doing so loses its memory of the details about the initial conditions. In this simulations the system goes through a universal scaling regime, where the gluon distribution function behaves as 
\begin{equation}
f_g(\tau; \,p_\perp, p_z)= \frac{1}{\alpha_s}\,\tau^{\alpha}\,f_S (p_\perp\,\tau^{\beta},  p_z\,\tau^{\gamma})\,.
\label{eq:scaling}
\end{equation} 

Here, $f_S$  is a time independent function, whose shape is given by non-perturbative physics of the theory. The exponents $\alpha=-2/3$, $\beta=0$ and $\gamma=-1/3$ confirm the parametrical descriptions of the BMSS scenario \cite{Berges:2013lsa}, thus identifying its approach to thermalization as the correct description of the expanding Glasma. 

In this paper, I use the $2\leftrightarrow 2$ kinetic photon rate to calculate the $p_\perp$ resolved spectra following the assumptions for the estimates of ref. \cite{Berges:2017eom}. This is done in the context of the \textit{bottom-up} thermalization scenario. The rate is further simplified using a small-angle approximation.
For this calculation, the momentum dependent non-thermal distribution of quarks is needed, which is sampled via the hard $g\rightarrow q\bar{q}$ approximation. This means that $f_q \sim \alpha_s f_g$, where $f_g$  is taken to be the non-equilibrium scaling solution from eq. (\ref{eq:scaling}). The non-equilibrium rates are enhanced using a bremsbremstrahlung ansatz analogous to the complete leading order (LO) thermal rate \cite{Arnold:2003rq}. The different contributions from the Glasma, as well as the later thermal stage, will be compared, to establish their relative dominance in this model. Finally, I will fix the model's parameters phenomenologically to present a qualitative comparison with data. 

This paper is organized as follows. In section \ref{sec:frame} the reader can find a small account of the kinetic framework used, as well as the low-$p_\perp$ enhancement ansatz used for the non-equilibrium case. In section  \ref{sec:NONEQ} I summarize the BMSS thermalization scenario and present the $2\leftrightarrow 2$ leading log (LL) results for the Glasma in this context. In sec. \ref{sec:parfix}, I will review the parameter fixing of Ref. \cite{Berges:2017eom} and apply it to the $p_\perp$ resolved spectra. The main body of the results achieved in this work is given in sec. \ref{sec:results}. Here, a comparison of the different contributions to direct photons will be presented for both the LL and the LO case. Furthermore, I will also compare photon production in the BMSS scenario with its early thermalization counterpart. Finally, a qualitative comparison to ALICE and PHENIX data will be presented.  This will be followed by conclusions and outlook. In the Appendices \ref{sec:Anal} and \ref{sec:TH} I present the derivation of the pre-equilibrium and thermal yields, respectively. 

\section{Approximate kinetic description}
\label{sec:frame}
Following Ref.~\cite{Berges:2017eom}, the photon rate for a thermalizing colored medium will be calculated using a kinetic description. The emission rate of an on-shell photon with three-momentum $
\boldsymbol{p}=(p_x,p_y,p_z)$ at a space-time point $
X=(t, x,y,z)$ from two-to-two scatterings is given generally by~\cite{Kapusta:1992gv,Baier:1991em}

\begin{equation}
\begin{aligned}
E\frac{\mathrm{d}N }{\mathrm{d}^4X \mathrm{d}^3p }\, =\, & \frac{1}{2\,(2\,\pi)^{12}}\int \frac{\mathrm{d}^3\,p_3}{2E_3} \frac{\mathrm{d}^3\,p_2}{2E_2}  \frac{\mathrm{d}^3\,p_1}{2E_1}\,|\mathcal{M}|^2\\
&\quad \times\, (2\,\pi)^4\, \delta^4(P_1+P_2-P_3-P) \\
& \quad\times\, f_1(p_1)\,f_2(p_2) \left[1\pm f_3(p_3)\right]\,\,,\,
\end{aligned}
\label{eq:kinetic}
\end{equation}
with $P_i=(E_i,\boldsymbol{p}_i)$, $i=1,2,3$.
The total squared amplitude $|\mathcal{M}|^2$ is understood as summed over spins, colors and flavors of all in- and outgoing particles. For massless quarks, the annihilation of a quark-antiquark pair into a photon and a gluon yields the squared amplitude
\begin{equation}
|\mathcal{M}_{\mathrm{anni}}|^2 =\frac{160}{9}\,16\pi^2\,\alpha\,\alpha_s\, \frac{u^2+\bar{t}^2}{u\,\bar{t}},
\end{equation}
with the strong interaction coupling $\alpha_s$ and the electromagnetic coupling $\alpha$. 
The squared amplitude for mixed Compton scattering, where a gluon kicks a (anti)quark producing a photon is  
\begin{equation}
|\mathcal{M}_{\mathrm{Comp}}|^2 =\frac{320}{9}\,16\pi^2\,\alpha\,\alpha_s\, \frac{u^2+s^2}{-u\,s}.
\end{equation}
These are given in terms of the Mandelstam variables \mbox{$s=(p_1 +p_2)^2$}, $\bar{t}=(p_1-p)^2$, and $u=(p_3-p_1)^2$.

To estimate the scattering rates, I will consider these two processes. Furthermore, the production of photons may be simplified using the small-angle approximation~\cite{Blaizot:2014jna,Tanji:2017suk}. In this case, one finds the rate 
\begin{equation}
E\frac{\mathrm{d}N }{\mathrm{d}^4X \mathrm{d}^3p }= \frac{40}{9\pi^2} \, \alpha \, \alpha_S \,\mathcal{L} \,  f_q (\mathbf{p})\, I_g \, ,
\label{eq:smallangle}
\end{equation}
where $f_q$ is the quark distribution. The $I_{g}$ integral is given by 
\begin{equation}
I_{g,q}= \int\frac{\mathrm{d}^3p}{(2\pi)^3}\frac{1}{p} f_{g,q}(\boldsymbol{p})\,.
\label{eq:igq}
\end{equation}
The Coulomb logarithm $\mathcal{L}$ in (\ref{eq:smallangle}) serves as a regulator and quantifies the ratio between the infrared and hard scales of the system, 
\begin{equation}
\mathcal{L}= \log\left(\frac{\Lambda_{UV}}{\Lambda_{IR}}\right) \, .
\label{eq:CL}
\end{equation}
As an example, one can take an isotropic thermal medium, where the hard scale $\Lambda_{UV}$ is given by the temperature $T$, while the infrared scale for the kinetic description is defined by the screening mass, $m_D \sim g T$, with $\alpha_s \equiv g^2/(4\pi)$. More generally, the Debye screening mass squared of the medium is estimated by $m_D^2 = 4 g^2 (N_c I_g + N_f I_q)$.

The rate (\ref{eq:smallangle}) still neglects the resummation of multiple interactions with the medium that would contribute at leading order in $\alpha$ and $\alpha_s$~\cite{Arnold:2001ba}. In thermal equilibrium the naive rate (\ref{eq:smallangle}) differs from the full leading-order result by about a factor of two in the relevant photon momentum range. For simplicity, and to also effectively take this into account, I adopt the prescription of Refs.~\cite{Berges:2017eom,Kapusta:1992gv} to replace the Coulomb logarithm in (\ref{eq:CL}) by 
\begin{equation}
\mathcal{L} \longrightarrow 2 \log\left(1+ 2.912/g^2\right) \, 
\label{eq:Log}
\end{equation}
to match the leading log (LL) thermal result. This will be employed for all the small-angle estimates shown in this work.

\subsection{Bremsstrahlung Ansatz}
\label{sec:IRE}

The small angle approximation gives the correct limit for thermal radiation at the LL level, once the coulomb logarithm has been identified as in eq. (\ref{eq:Log}). Nonetheless, it has been shown in ref. \cite{Arnold:2001ms} that in a thermal medium, near-collinear bremsstrahlung dominates the rates for photon energies of $p\lesssim 2\,T$, while at intermediate photon momenta, $2\,T\lesssim p \lesssim 10\, T$, the $2\leftrightarrow 2 $ contribution is comparable to the near-collinear ones. 
One would expect this simple result to be the case in the non-equilibrium setting of the Glasma, with one general caveat. In the Glasma, the characteristic momentum scale is given by $Q_s$, making the near-collinear contributions during the early stages dominant at $p\lesssim 2\,Q_s$ which for RHIC and LHC energies covers most of the kinematic window at which excess has been observed ($0.5-3\,\mathrm{GeV}$).

Following this argument, the bremmstrahlung contribution has to be included, and we will do so by changing the \textit{total constant under the log} in eq. (\ref{eq:smallangle}) where the temperature has to be substituted for the characteristic scale of the Glasma. Following the result from \cite{Arnold:2001ms} the leading order (LO) thermal rate can be expressed as

\begin{equation}
E\frac{d\,N }{d^4\,X\,d^3\,p }=A(p)\,\nu\left(\frac{E}{T}\right)\,,
\end{equation}
where  
\begin{equation}
A(p)= 2\alpha d_F \left[ \sum_c q_c^2\right]\,m^2_D\,  f_{q,eq}\left(\frac{E}{T}\right)\,,
\end{equation}
with the thermal screening mass $m^2_D=C_F \,g^2_s\, T^2 /4$ and the Casimir operators   $d_F=3$ and $C_F=4/3$ for $SU(3)$. The function $\nu$ is here the total constant under the log, and can be represented by different functions for both the LL and LO cases, 

\begin{align*}
&LL: \nu\left(\frac{E}{T}\right) \rightarrow \nu_{LL}\left(\frac{E}{T}\right) \rightarrow \mathcal{L}\\
&LO: \nu\left(\frac{E}{T}\right) \rightarrow \nu_{LL}\left(\frac{E}{T}\right) + C_{bremss}\left(\frac{E}{T}\right)+C_{anni}\left(\frac{E}{T}\right)\,.
\end{align*}

The explicit forms of $C_{bremss}$ and $C_{anni}$ can be found in ref. \cite{Arnold:2001ba}.
In the non-equilibrium case, one can expand the former results by using the same LO function, while changing the temperature dependence for the appropriate characteristic scale of the system,  $Q_s$. Using this change, one can write down the function as follows 
\begin{align*}
&LL: \nu\left(\frac{E}{Q_s}\right)\rightarrow \nu_{LL}\left(\frac{E}{Q_s}\right)  \rightarrow \mathcal{L}\,\\
&LO: \nu\left(\frac{E}{Q_s}\right) \rightarrow \nu_{LL}\left(\frac{E}{Q_s}\right) + C_{bremss}\left(\frac{E}{Q_s}\right)+C_{anni}\left(\frac{E}{Q_s}\right)\,.
\end{align*}

In the next sections, both the $2\leftrightarrow 2$ and the LO ansatz results will be presented to allow an appropriate comparison with the estimates presented by Berges et al~\cite{Berges:2017eom}, as well as to show a better case scenario for photons coming from the BMSS scenario. For this, the non-equilibrium rates of Stages (i) and (ii) will be computed via eq. (\ref{eq:smallangle}) but would receive near-collinear enhancements thanks to the substitution from above.
In the case of the third stage, the rate is already given by the LL thermal rate, which means it can be just upgraded to the LO thermal rate, as it is parametrized in ref. \cite{Arnold:2001ms}, for the appropriate space-time dependence of the temperature.
\section{Non-equilibrium Sources}
\label{sec:NONEQ}
The main assumption of the \textit{bottom-up} scenario~\cite{Baier:2000sb}, and also this work, is that gluonic saturation physics takes place for energies at LHC and RHIC. In this scheme, the non-thermal colored medium undergoes  three stages in its path for thermalization, which can be parametrically separated  as
\begin{align*}
&\mathrm{(i)}  \,\,\,1\ll Q_s\,\tau\ll \alpha_s^{-3/2}\\
&\mathrm{(ii)}\,\,\, \alpha_s^{-3/2}\ll Q_s\,\tau\ll \alpha_s^{-5/2}\\
&\mathrm{(iii)} \,\,\, \alpha_s^{-5/2}\ll Q_s\,\tau\ll \alpha_s^{-13/5}
\end{align*}

In the beginning of the collision, at $\tau<Q_s^{-1}$ the physics of the glue is highly  non-perturbative, and the states can be characterized by very non-linear macroscopic fields. During this stage, instabilities highly populate modes with $p_\perp\lesssim Q_s$ \cite{Romatschke:2005pm}. After these modes have been occupied, the system is completely dominated by hard modes, for which  $p_\perp \sim Q_s$. 
These modes are approximately conserved, but due to Bjorken expansion their number density is diluted as $n_{h}\sim Q_s^3/(Q_s\,\tau)$. 
During this stage, gluons interact via $2\leftrightarrow 2$ hard scatterings, with a very small momentum exchange. This produces a broadening, or melting, of the distribution of longitudinal momentum $p_z\sim Q_s\,(Q_s\,\tau)^{-1/3}$. The produced effect is a decrease of the typical occupation number as $f_g \sim \alpha_s^{-1}\,(Q_s\,\tau)^{-2/3}$. 

The second stage starts when the typical occupation $f_g$ falls below unity. This happens parametrically at $Q_s\,\tau \sim \alpha_s^{-3/2}$.  During this stage, the number of soft gluons rise rapidly via collinear splitting.  Nonetheless, hard gluons still dominate the total number, with their number densities given by 
\begin{equation}
n_{h}\sim\frac{Q_s^3}{Q_s\,\tau}    \quad\mathrm{and}\quad n_{s}\sim\frac{\alpha_s^{1/4} \,Q_S^3}{(Q_S\,\tau)^{1/2}}\,.
\end{equation}  

In this stage, soft gluons possess a typical momentum of $p_{soft}\sim \alpha^{1/2}\,Q_s$. It can be seen that $n_{soft}/p_{soft}\gg n_{hard}/p_{hard}\sim n_{hard}/Q_s$. This makes the Debye integral peaked strongly around the soft sector, from which $m_D$ can be found to be $m_D^2\sim \alpha_s \,n_{soft}/p_{soft} \sim\alpha_s^{3/4} \,Q_s^2 \,(Q_s\,\tau)^{-1/2}\,$. The typical longitudinal momentum of hard gluon stops decreasing in (ii) and converges to $p_z\sim \alpha_s\, Q_s$, which means that the anisotropy of the system saturates at finite value.

The thermalization stage (iii) starts around the time $Q_s\,\tau\sim \alpha^{-5/2}$, where $n_{h}$ and $n_{s}$ become comparable, while $p_{soft}\ll p_{hard}$. This signals that soft modes are now dominant both in number densities and in the screening mass. Soft gluons thermalize very fast via  $2\leftrightarrow 2$ soft scatterings, and act as a thermal bath to which the hard sector looses energy via mini-jet quenching \cite{Blaizot:2013vha}. Since hard gluons act as a source of energy to the bath, the temperature rises with $T=c_T\, \alpha^3_s \, Q_s^2\, \tau$, to finally achieve full thermalization of the medium at 

\begin{equation}
\tau_{th}\sim c_{eq}\,\alpha^{-13/5} Q_s^{-1}   \quad\mathrm{and}\quad T_{th}\sim c_{T}\,c_{eq}\,\alpha^{2/5} Q_s\,.
\end{equation}  

Using this model to estimate the evolution of the Glasma, I can use these results to calculate the photon spectra produced on the road to thermalization. 

In the following, I use the variables $\tau=\sqrt{t^2 -z^2}$, $\eta=\mathrm{arctanh}\left(z/t\right)$ and $y= \mathrm{arctanh} (p_z/E)$. For the transverse plane, I employ a polar parametrization, $p_x=p_\perp \cos\phi $ and $p_y=p_\perp \sin\phi $ in terms of the transverse momentum $p_\perp$, longitudinal momentum $p_z$ and azimuthal angle $\phi$. The photon multiplicities will be obtained by integrating the eq. (\ref{eq:smallangle}) over the four-volume of the evolution for each stage, using $\mathrm{d}^4X= \tau \mathrm{d}\tau \mathrm{d}\eta d^2 x_\perp$ and $\mathrm{d}^3p/E = \mathrm{d}y\mathrm{d}^2p_\perp$. 

\subsection{Glasma, Stage I  }

In the first stage of the evolution of the Glasma, quarks are taken to inherit their properties from the gauge sector via hard gluon splitting. This means that $f_q\sim \alpha_s\, f_g$, where $f_g$ is the gluon distribution found in classical statistical simulations. A small caveat has to be noted here regarding the quantum statistics of this function. This approximation is only valid while $\alpha_s \,f_g\ll 1$ which  will be the case for realistic parameters.
The distribution exhibits self-similarity, and during the scaling regime,  dynamics is given by a time-independent function $f_S$, from which one gets the gluon distribution via the relation 
\begin{equation}
f_g(\tau; \,p_\perp, p_z)= \frac{1}{\alpha_S}(Q\tau)^{-2/3}\,f_S (p_\perp,  p_z\,(Q\tau)^{1/3})\,.
\label{eq:glue1}
\end{equation}

This scaling solution, $f_S$, was found in numerical studies for Bjorken expanding lattices \cite{Berges:2013eia,Berges:2013fga,Berges:2013lsa}, and it is given by the form 
\begin{equation}
f_S(p_\perp, p_z)=f_0 \frac{Q}{p_\perp}\, \exp\left[{-\frac{1}{2}\frac{p_z^2}{\sigma^2_0}}\right]\, W_r[p_\perp-Q_s]\,,
\label{eq:glueS}
\end{equation}
which was fitted from the results of ref. \cite{Berges:2013lsa},
where I define  $W_r[p_\perp-Q_s]$ as the function that guarantees the suppression of the distribution function around $p_\perp = Q_s$, as observed by simulations in \cite{Berges:2013eia}. The parametrization taken from the fits for $W_r[p_\perp-Q_s]$ is as follows,
\begin{equation}
\WR = \theta(Q_s - p_\perp) +  \theta(p_\perp-Q_s ) \, 
e^{-\frac{1}{2}(\frac{p_\perp-Q_s }{r\,Q_s})^2}\,,
\end{equation}
where $r$ is a free parameter  of $\mathcal{O}(1)$ that allows the correct suppression at higher momenta. During this stage, the gluon occupancy is dominated by hard gluons, which are approximately conserved, up to the expansion dilution factor $\tau^{-1}$. This behaviour determines the time dependence of $I_g$, which can also be found by simply using the scaling properties in eq. (\ref{eq:glue1}). An overall normalization constant, $\kappa_g$, is used as a  proportionality constant. This constant was found in ref. \cite{Berges:2017eom} to be given $\kappa_g = c/(2\,N_c)$ where $c=1.1$ is the gluon liberation coefficient \cite{Lappi:2007ku}. Having this together, the gluon integral as 

\begin{equation}
I_g(\tau)=\frac{Q^2_s}{4\,\pi^2\alpha_s}\frac{\kappa_g}{(Q_s\,\tau)}\,.
\end{equation}

Using the aforementioned ingredients, the rate for the first stage is

\begin{equation}
E\frac{\mathrm{d}N }{\mathrm{d}^4\,x \mathrm{d}^3\, p }= \frac{10}{9\pi^4} \, \alpha \,\mathcal{L} \, \kappa_g\,\frac{Q^2_s}{(Q_s\,\tau)}  f_q (\mathbf{p})\,.
\label{noneq1}
\end{equation}

For the computation of the photon multiplicity of the first stage, I assume that the photon momentum $p$ is on shell, which means $p_0= p_\perp\cosh(y-\eta)$, as well as  $p_z = p_\perp\sinh(y-\eta)$. Integrating over the spatial variables, I get the expression 
\begin{equation}
\begin{aligned}
\frac{1}{S_\perp}\frac{\mathrm{d}N_\gamma }{\mathrm{d}^2 p_\perp\mathrm{d}y} =  \sqrt{\frac{\pi}{2}}\frac{20}{9\pi^4} \alpha_e\,\kappa_g & \,f_0\, \frac{\sigma_0}{Q_s} \,\log\left(\frac{\tau_1}{\tau_0}\right) \\ 
&\times\, \frac{Q^2_s}{p^2_\perp}\, W_r[p_\perp-Q_s] \,,
\end{aligned}
\label{eq:GL11PDAP}
\end{equation}
using the limit of $\sigma_0/Q_s \rightarrow 0$ \footnote{The details of the derivation of the Glasma yields can be found in Appendix A.  The derivation of the thermal QGP yield can be found in Appendix B. }.
The initial and final times are then substituted for $ \tau_0 = c_0\, Q_s^{-1} $ and $ \tau_1= c_1\, \alpha_s^{-3/2}\,Q_s^{-1}  $, where $c_0$ and $c_1$ are unknown proportionality constants of order $\mathcal{O}(1)$. As it was stated before, a precise description of the Glasma will include such coefficients, but it is out of the scope of this work. In this model, these coefficients will be set to unity. This is in fact, supported by slow, logarithmic dependence of the coefficient ratio $c_1/c_0$  in eq. (\ref{eq:GL11PDAP}).
  \begin{figure}[t]
  \centering
  \includegraphics[width=0.47\textwidth]{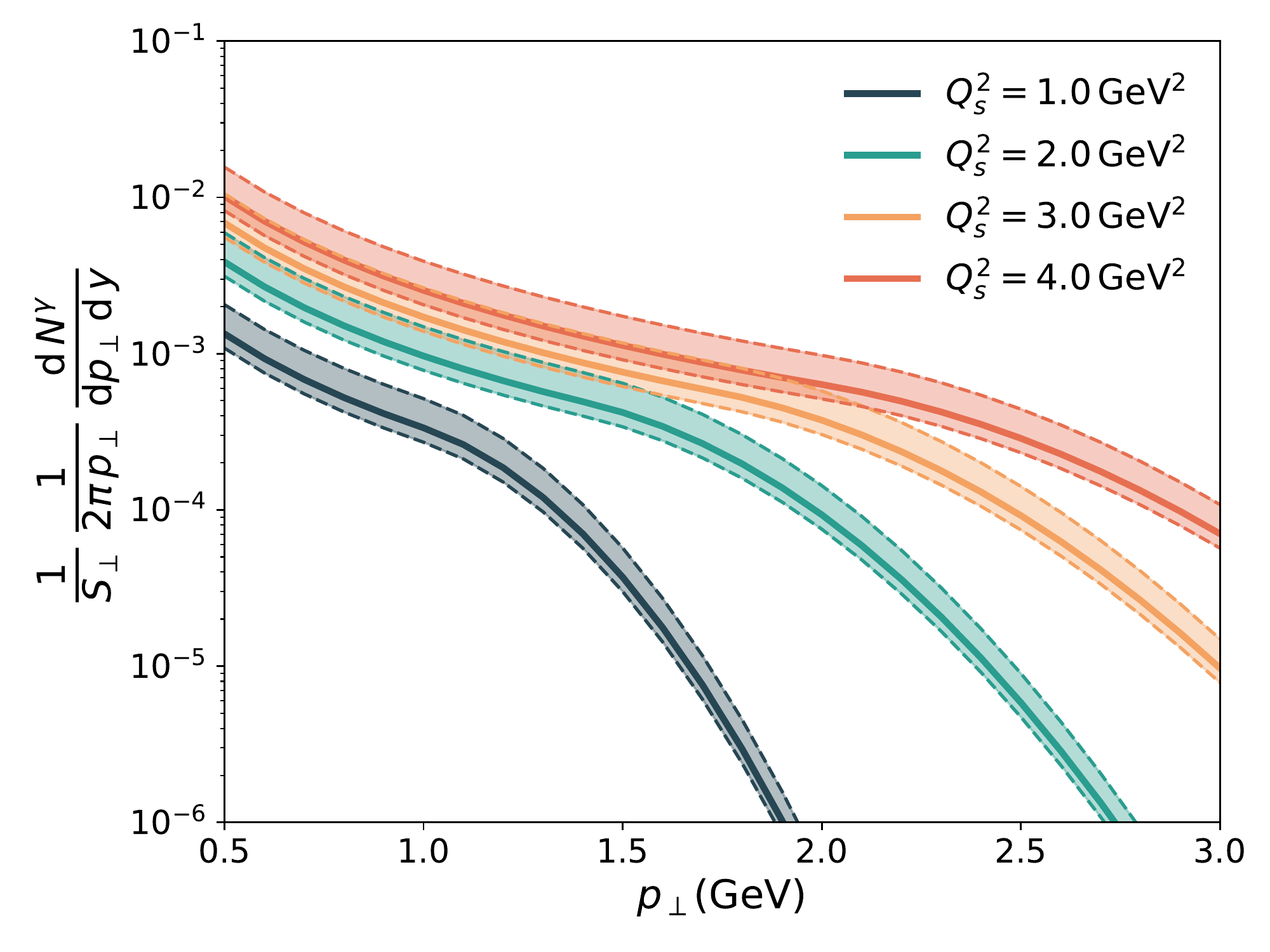}
  \caption{Photon invariant yield from the first stage of the Glasma for different values of $Q_s^2$. The error bands correspond to a factor of $2$ variations of the anisotropy parameter,$\sigma_0$.}
\label{fig:glass1}
\end{figure}
The expansion in eq. \eqref{eq:GL11PDAP} is safe as long as one is interested in the $p_\perp>\sigma_0 $ portion of the spectrum. Experimentally, this is the case, as we will be interested in the kinematic window $p_\perp \geq 1 \mathrm{GeV}$,  and given that in RHIC and LHC, the characteristic saturation scales are thought to be  $Q_s\gtrsim 1 \mathrm{GeV}$ for heavy ions, we can see that the interesting momenta for this setup satisfy this condition, for $\sigma_0\lesssim 0.1\,Q_s $. Finally one can further integrate the multiplicities over the radial momenta to find the multiplicity of photons per unit rapidity. In the totally anisotropic limit, i.e. vanishing $\sigma_0/Q_s$, the results can be simplified to 

\begin{equation}
\begin{aligned}
\left.\frac{1}{S_\perp}\frac{\mathrm{d}N_\gamma }{\mathrm{d}y}\right|_{p_\perp<Q_s} =& \sqrt{ \frac{\pi}{2}}\frac{20}{9\,\pi^3} \alpha_e\,\kappa_g\,f_0\, \sigma_0\, Q_s \, \\ 
&\times\log(\alpha_s^{-3/2}) \log
   \left(\frac{Q_s^2\, }{\sigma_0^2}\right) \,.
   \end{aligned}
  \label{eq:TotalYieldI} 
\end{equation}

Which is identical to the result in ref. \cite{Berges:2017eom}  after performing matching the normalisation of the distribution function , $f_0$ in terms of the anisotropy parameter,  $\sigma_0/Q_s$,

\begin{equation}
f_0\,\frac{\sigma_0}{Q_s} = \sqrt{\frac{2}{\pi}} \kappa_q \left[\log\left(\frac{Q_s^2}{\sigma_0^2}\right) \right]  ^{-1}\, , 
\label{eq:match}
\end{equation}

 I will use this matching  from now on this work. The rest of the yield can be found by integrating eq. (\ref{eq:GL11PDAP}) above the saturation scale $Q_s$, which gives the result 
\begin{equation}
\begin{aligned}
\left.\frac{1}{S_\perp}\frac{\mathrm{d}N_\gamma }{\mathrm{d}y}\right|_{p_\perp<Q_s} =& \sqrt{ \frac{\pi}{2}}\frac{40}{9\,\pi^3} \alpha_e\,\kappa_g\,\kappa_q\, Q_s^2 \, \chi_r \,\\ 
&\times\log(\alpha_s^{-3/2}) \left[\log\left(\frac{Q_s^2}{\sigma_0^2}\right) \right]  ^{-1} \,,
\end{aligned}
 \label{eq:GL1YAP}
\end{equation}
where 
\begin{equation}
 \chi_r = \frac{1}{2} e^{-\frac{1}{2 r^2}} \left(\pi  \text{erfi}\left(\frac{1}{\sqrt{2}
   r}\right)-\text{Ei}\left(\frac{1}{2 r^2}\right)\right)\,.
   \label{eq:chir}
\end{equation}
In the strict limit of full anisotropy, this contribution vanishes. For the set of parameters used in this work it will contribute to around $5\%$ of the total yield from the Glasma. From these results it can be seen that the total yield of photons of the first stage of the Glasma is basically insensitive to the fit parameters used in eq. (\ref{eq:glueS}). 

\subsection{Glasma, Stage 2 }

After a time $\tau\sim Q_s^{-1}\,\alpha_s^{-3/2}$, the typical gluon occupation drops below unity, and the rate should be revised. At this point, hard gluons still dominate the total number density, but soft modes take over the behaviour of the Debye mass. This change affects the time dependence of the $I_g$ integral, which behaves as $I_g\sim \alpha_s^{-1}\,m_D^2$ \cite{Baier:2000sb}. This leads to the expression

\begin{equation}
I_g(\tau)=\frac{\kappa_g\,\alpha_s^{-1/4}}{4\,\pi^2 \sqrt{c_1}}\frac{Q^2_s}{(Q_s\,\tau)^{1/2}}\,,
\label{eq:debye2}
\end{equation}
where the overall normalization of $I_g$ has been modified to match the expression of stage (i) at $\tau\sim Q_s^{-1}\,\alpha_s^{-3/2}$. The fermionic sector is always dominated by hard quarks, which can still be described by $f_q=\alpha_s\, f_g$, with $f_g$ as in eqs. (\ref{eq:glue1}) and (\ref{eq:glueS}). With those changes, the full rate for stage (ii) now is given by 
\begin{equation}
E\frac{\mathrm{d}N }{\mathrm{d}^4\,x \mathrm{d}^3\, p }= \frac{10}{9\pi^4} \, \alpha\alpha_s^{3/4} \,\mathcal{L} \, \frac{\kappa_g}{\sqrt{c_1}}\,\frac{Q^2_s}{(Q_s\,\tau)^{1/2}}  f_q (\mathbf{p})\,.
\end{equation}

The photon multiplicity for this stage can be found by again integrating over the full space-time volume and expanding to leading order in the anisotropy parameter, to get  
\begin{equation}
\begin{aligned}
\frac{1}{S_\perp}\frac{\mathrm{d}N^{(ii)}_\gamma }{\mathrm{d}^2 p_\perp\mathrm{d}y} = & \sqrt{2\,\pi}\frac{20}{9\pi^4} \alpha_e\,\kappa_g\,f_0\, \frac{\sigma_0}{Q_s}  \, \frac{Q^2_s}{p^2_\perp}\, \\ 
	& \times W_r[p_\perp-Q_s] \,\left(\sqrt{\frac{c_2}{c_1}}\alpha_s^{-1/2}-1\right)\,.
\end{aligned}
\label{eq:GL21PDAP}
\end{equation}

In the weak coupling limit, the system stays in (ii) a parametrically long time, which naturally leads to a higher yield than the stage (i). Here, as in the first stage, the dependence on the $c_2/c_1$ ratio is also slow and, since we expect these coefficients to be of order $\mathcal{O} (1)$, it will be taken to be unity. 

The known result for the total yield can be found by again expanding in $\sigma_0/Q_s$ up to lowest order. Applying the matching from eq. (\ref{eq:match}) I get

\begin{equation}
\begin{split}
\left.\frac{1}{S_\perp\,Q^{2}_S}\frac{\mathrm{d}N_\gamma }{\mathrm{d}y}\right|_{p_\perp<Q_s} = \sqrt{ \frac{\pi}{2}}\frac{40}{9\,\pi^3} &\alpha_e\,\mathcal{L}\, \kappa_g\, \kappa_q\\
&\times\,\left(\alpha_s^{-1/2}-1\right)\,.
\end{split}
\label{eq:S2TY}
\end{equation}

This result is identical to the one found the previous estimate \cite{Berges:2017eom}. The yield for photons with  $p_\perp> Q_s$  can be found by integrating eq. (\ref{eq:GL21PDAP}), 

\begin{equation}
\begin{split}
\left.\frac{1}{S_\perp\,Q^{2}_S}\frac{\mathrm{d}N_\gamma }{\mathrm{d}y}\right|_{p_\perp>Q_s} =  \sqrt{\frac{\pi}{2}}\frac{80}{9\pi^3}& \alpha_e\,\mathcal{L}\,\kappa_g  \,f_0\, \frac{\sigma_0}{Q_s} \\ 
&\times\chi_r  \,\left(\alpha_s^{-1/2}-1\right)\,,
\end{split}
\label{eq:GL2Y}
\end{equation}
where  $\chi_r$ was defined in eq. (\ref{eq:chir}). Once again, it can be observed that the total yield exhibits independence to the fit parameters, just as in stage $(i)$.
\begin{figure}[t]
  \centering
  \includegraphics[width=0.44\textwidth]{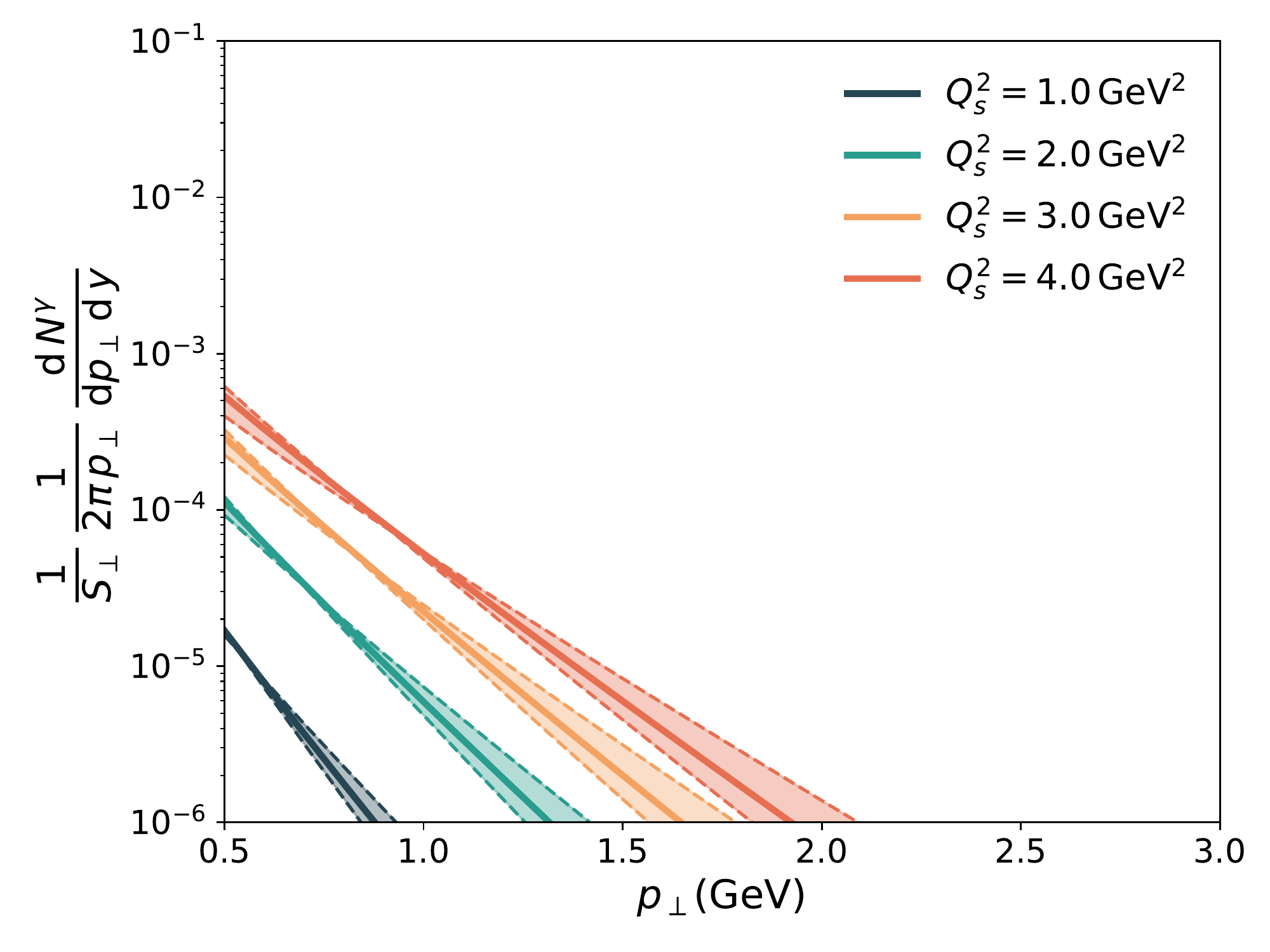}
  \caption{Photon invariant yield from the first stage of the Glasma for different values of $Q_s^2$. The error bands correspond to a factor of $2$ variations of the anisotropy parameter,$\sigma_0$.}
\label{fig:glass3}
\end{figure}
\subsection{Glasma, Stage 3 }
 
During this stage, soft gluons take over the number density. The beginning of this stage is at time parametrically larger than the relaxation time, $\tau_2>\tau_\mathrm{rel}$, which means that soft gluons modes  have thermalized already. Therefore, the distribution can be calculated using the rapidity integrated thermal rate (see Appendix \ref{sec:TH}), which can be further integrated in time to find the invariant yield, 

\small
\begin{equation}
\frac{1}{S_\perp}\frac{\mathrm{d}N_\gamma }{\mathrm{d}^2 p_\perp\mathrm{d}y}  =  \, \frac{5}{9}\,\frac{\aem\,\as}{\pi^2} \, \mathcal{L}\int_{\tau_{2}}^{\tau_{th}} \rmd\tau\,\tau\, T^2(\tau)K_0(\PTS{}/T(\tau))  \,.
\end{equation}
\normalsize 

In this stage, the temperature of the soft gluon bath increases linearly in time, and can be parametrized as
\begin{equation}
T=c_T\, \alpha_s^3\,Q_s^2\, \tau
\end{equation}
Substituting the time parameter $\tau$ in the thermal rate integral, I get
\begin{equation}
\frac{1}{S_\perp}\frac{\mathrm{d}N_\gamma }{\mathrm{d}^2 p_\perp\mathrm{d}y}  =  \, \frac{5}{9}\,\frac{\aem\,\as^{-5}}{\pi^2 c_T^2  }\, \frac{\mathcal{L}}{Q^{4}} \int_{T_2}^{T_{th}} \rmd T\, T^3\, K_0(\PTS{}/T)  \,.
\end{equation}

Once  again it is important to notice that the parametrisation used in this equation works only for values of the transverse momentum such that $p_\perp \gtrsim T$. However, for the kinematic window of interest, $p_\perp \geq 1\,\mathrm{GeV}$, it is also parametrically satisfied that $p_\perp >  T_{th}$, which  means that the rate can be approximated using the asymptotic form 
\begin{equation}
K_0(x) \sim \sqrt{\frac{\pi}{2\,x}}\,e^{-x} \quad\mathrm{for}\quad x \gg 1\, .
\end{equation} 

Throughout this stage the temperature rises, which makes the assumption safer, since $T<T_{th}< p_\perp$. In the next section I will show that for the photon kinematic window and for realistic parameters this is always the case. Using this approximation, the integral for the photon invariant yield in the third stage can be found to be 
\begin{widetext}
\[
\frac{1}{S_\perp}\frac{\mathrm{d}N_\gamma }{\mathrm{d}^2 p_\perp\mathrm{d}y}  =\sqrt{\frac{\pi}{2}} \, \frac{5}{9}\,\frac{\aem\,\as^{-5}}{\pi^2 c_T^2  } \frac{p_\perp^4}{Q^{4}} \frac{32}{945} \, \mathcal{L} \left( \sqrt{\pi } 
   \text{erf}\left(\sqrt{\frac{p_\perp}{T}}\right)\right.   
  \left. +\sqrt{\frac{T}{p_\perp}} e^{-\frac{p}{T}} \left(1
   - \frac{1}{2}\frac{T}{p_\perp}+\frac{3}{4} \frac{T^2}{p^2_\perp}
   -\left.\frac{15}{8} \frac{T^3}{p_\perp^3}+\frac{105}{16} \frac{T^4}{p_\perp^4}\right)\right)\right|_{T_2}^{T_{th}}\,.
\]
\end{widetext}

\section{Parametrical fixing and comparison}
\label{sec:parfix}
The calculations presented above depend on a collection of parameters which describe the non-equilibrium dynamics of the BMSS scenario. To reduce the parameter phase-space, one can use a phenomenological matching to pin down some of these parameters. The first coefficient to fix is the semi-hard scale, $Q_s$, from which all the BMSS dynamics is dependent. It can be found in terms of the energy and $N_{part}$ using IP-Glasma model \cite{Schenke:2012wb,Schenke:2012fw,Schenke:2013dpa,TribedyPrivate}. The IP-Glasma model combines the IP-Sat model  \cite{Kowalski:2003hm, Rezaeian:2012ji}  with the MC-Glauber model dependence on the geometry of the system  \cite{Miller:2007ri}.  Since only the mixed quantity $\langle S_\perp\, Q^2_s \rangle\equiv \int d^2\boldsymbol{x}_\perp Q_s(\boldsymbol{x}_\perp, \sqrt{s})$ can be calculated in terms of $N_{part}$, I will approximate the transverse average $Q_s(x,N_{part})$ by using 
\begin{equation}
Q^2_s(x,N_{part})=\frac{ \langle S_\perp\, Q^2_s \rangle}{\langle S_\perp\rangle}\,,
\end{equation}
where $\langle S_\perp\rangle$ is calculated using the Glauber model. 

I have adjusted the overall normalization to a reference $Q_s^2 = 2\,\mathrm{GeV} $ for the highest centrality class, $0-5\%$, with $N_{part}=353$. This can be done because we lack a first-principles determination of the hard scale in the Glasma, which makes it possible to vary the definition while keeping in mind it should be a \textit{semihard} scale. 

\begin{table*}[]
\begin{tabular}{cccccc}
              & Centrality   & $N_p$   & $S_\perp\, (\mathrm{fm}^2 )$    & $Q_s^2\, (\mathrm{GeV}^2 )$ &   $c_{eq}\, c_T^{3/4}$ \\ \hline
\multirow{3}{*}{\begin{tabular}[c]{@{}c@{}}PHENIX\\ 200GeV\end{tabular}} 
			 & 0-5\%     &$\,\,353.0 \pm 10.0\,\,$ &$\,\,143.3 \pm 2.7 \,\,$&$\,\,2.00 \pm 0.04\,\,$& $\,\,0.306 \pm 0.005\,\,$ \\
              & 0-20\%   &$277.5 \pm 6.5$   &  $122.1 \pm 1.9$&  $ 1.67\pm 0.02$&     $0.318 \pm 0.006$                \\
              & 20-40\% &$135.5 \pm 7.0$   &  $75.7 \pm 2.6$ &  $ 1.12\pm 0.04$&     $0.343 \pm 0.008$               \\ \hline
\multirow{3}{*}{\begin{tabular}[c]{@{}l@{}}ALICE \\ 2.76 TeV\end{tabular}}
		     & 0-5\%  	  &$383.5 \pm 1.9$  &  $155.8 \pm 0.5$ & $ 3.60\pm 0.01 $&$ 0.278\pm 0.004$ \\
              & 0-20\%   & $307.2 \pm 2.6$  &  $135.6 \pm 0.7 $ &  $ 2.90\pm 0.02$ &   $0.289 \pm 0.004 $ \\
              & 20-40\% & $160.3 \pm 2.7$    &  $87.1 \pm  1.0$   &    $ 1.81\pm 0.02$   &      $0.321 \pm 0.006$          
\end{tabular}
\caption{Relevant parameter fixing for diverse centrality clasess in RHIC and LHC . For details on the fixing process see text and ref. \cite{Berges:2017eom}}
\label{tab:params}
\end{table*}

  The next step is to relate the coefficients $c_{eq}$ and $c_T$, used in sec. \ref{sec:NONEQ}  and appendix  \ref{sec:TH} to measured quantities. For this the entropy per unit rapidity of QGP was matched to the entropy of produced charged hadrons. For an ideally expanding fluid, entropy is conserved, and one can perform the matching at thermalization time and temperature. The calculation is given in ref. \cite{Berges:2017eom}, and results in the expression 
 \begin{equation}
c_{eq}\,c_T^{3/4}=\left[ \frac{45}{148\pi^2}k_{S/N}\alpha^{7/5}\frac{N_{part}}{Q^2_s\,S_\perp}\frac{2}{N_{part}}\frac{\mathrm{d}N_{ch}}{\mathrm{d}y}\right]^{1/4}\,.
\label{eq:fixinfctceq}
\end{equation} 

Here, the experimental input is the multiplicity of charged hadrons per unit rapidity, $\mathrm{d}N_{ch}/\mathrm{d}y$,   in terms of $N_{part}$, given in ref. \cite{Aamodt:2010cz,Adler:2004zn,Abelev:2008ab}. The parameter $k_{S/N}$, is the proportionality constant that links total entropy of the hadronic phase with the measured multiplicity of charged hadrons. I will adopt the value $k_{S/N}= 7.2$  which has been extracted from particle spectra and particle interferometry \cite{Pal:2003rz,Gubser:2008pc}. I have adopted a running coupling parametrisation given by 
\begin{equation}
\alpha_s (Q_s) = \frac{12\pi}{(33-2N_f)\log\left(\frac{Q_s^2}{\Lambda_{QCD}^2}\right)}\,,
\end{equation} 
where I take $\Lambda_{QCD}=0.2\,\mathrm{GeV}$, and the number of in-medium quarks is taken to be $N_f=3$. Using this data I get the phenomenologically interesting parameters listed in table \ref{tab:params}, where $Q_s$, $S_\perp$, and $c_{eq}\,c_T^{3/4}$ are given for different centrality classes, that is $N_{part}$, for RHIC and LHC energies. The quantities $c_{eq}$ and $c_T$ can only be fixed together, as specified in eq. (\ref{eq:fixinfctceq}). However, the thermalization temperature coefficient was found in ref. \cite{Baier:2002qv} up to logarithmic accuracy to be $c_T = 0.18$, and will vary it over an overall factor of $2$. 

An important caveat is that, using the expressions for $\tau_{th}$ and $T_{th}$, as well for the temperature evolution in a Bjorken expansion, one can find also a phenomenological matching for $\tau_c$, which is given by 
\begin{equation}
\tau_c= \frac{45}{74\pi^2}k_{S/N}\frac{1}{S_\perp}\frac{\mathrm{d}N_{ch}}{\mathrm{d}\eta}\frac{1}{T_c}\,.
\end{equation}

This renders $\tau_c$ insensitive to $Q_s$ and $\alpha_s$, while dependent uniquely on the number of participants in the collisions, $N_{part}$. Because, for $N_{part}\lesssim150$ I find that $\tau_c\lesssim\tau_{eq}$ I will only take on account events which lie inside the $0-20\%$ centrality range. 

\section{Results}
\label{sec:results}
In the following, the comparisons will be made by including all the sources for direct photons. That is, apart from the BMSS and QGP radiation, one should include also thermally  produced photons from a hadronic gas (HG) after the hadronization of the QGP phase. Photons produced by hard scattering and annihilation of the participating partons, referred normally as \textit{prompt photons} have to be also included. The total direct invariant yield is then given by 

\begin{equation}
\frac{\mathrm{d}N }{\mathrm{d}^2 p_\perp\mathrm{d}y}= T_{AA}\, \frac{\mathrm{d}\sigma^{pp} }{\mathrm{d}^2 p_\perp\mathrm{d}y} + K^\gamma \left[\frac{\mathrm{d}N^{gl }}{\mathrm{d}^2 p_\perp\mathrm{d}y} +\frac{\mathrm{d}N^{th} }{\mathrm{d}^2 p_\perp\mathrm{d}y}\right]\,.
\label{eq:totaldirect}
\end{equation}
where the hadronic  the $\sigma^{pp}$ label stands for prompt photon (pQCD) cross section \cite{VogelsangPrivate}, which will be scaled by a centrality dependent factor, $T_{AA},$ with $A= \mathrm{Au} ,\,\mathrm{Pb}$. This can be calculated directly from the Glauber model. For the extension of the scaled pQCD  to lower $p_\perp$ values, which is needed to sum this contribution to the in-medium spectra, I have used the following functional form, 
\begin{equation}
 \frac{\mathrm{d}\sigma^{pp} }{\mathrm{d}^2 p_\perp\mathrm{d}y} =  A_{pp}\,\left(1+\frac{p^2_\perp}{P_0}\right)^{-n}\,,
\end{equation}
which was used by PHENIX to fit the $p+p$ results in \cite{Adare:2014fwh}.
In eq. (\ref{eq:totaldirect}), hadronic bremsstrahlung of photons is included as a thermal contribution, and summed over the QGP rate. In this work I included meson and baryonic rates, the $\pi\,\pi $ bremsstrahlung rate, as well as reactions of the $\pi\,\rho\,\omega $  meson system. \cite{Turbide:2003si,  Heffernan:2014mla, Holt:2015cda}. This is calculated naively for a Bjorken expanding system by switching from the QGP to the HG rate at a freezout temperature of $154\,\mathrm{MeV} $. A mild variation of this parameter did not affect the result. 

To account for the Glasma radiation, I have used $\sigma_0/Q_s = 0.15$ where a variation of $50\%$ has been included as error bands. It can be seen parametrically from eqs. (\ref{eq:GL11PDAP}) and (\ref{eq:GL21PDAP}) that the dependence on this parameter is parametrically slow. The suppression parameter is chosen to be $r=0.35$ for a better fit to data. It is always important to note that even when both $r$ and $\sigma_0$ are parameters physically motivated by ab-initio calculations, the particular value is picked as free parameter with soft constraints. This is indeed a source of systematic error in the model, but solving this issue with better simulations is outside the scope of this paper. Besides the gluon fitting parameters, other sources of uncertainty come from the thermalization time and temperature. These quantities vary with the constants $c_T$  and $c_{eq}$. As mentioned before, the product $c_{eq}\,c_T^{3/4} $  has been fixed together using eq. (\ref{eq:fixinfctceq}), which still leaves freedom to vary $c_T$ around the value calculated in ref. \cite{Baier:2002qv}. This will have an effect on the thermal invariant yield per transverse area since it is given parametrically by  $\tau_{th}^2\,T_{th}^2$ parametrically,  which means that after fixing it still depends polynomially on $c_T^{-10/3}$. For changes of a factor of $2$ in $c_T$, that still amounts for an extra $\mathcal{O}(1)$ coefficient when one takes on account the changes in the temperature. The Glasma stages (i) and (ii) are also mildly sensitive to the coefficients $c_0$, $c_1$ and $c_2$, which can result in the rates picking up also extra order  $\mathcal{O}(1)$ factors. Nevertheless, such fine tuning of the transition times is not the objective of this discussion.
 \begin{figure}[t!]
\centering
  \includegraphics[width=0.47\textwidth]{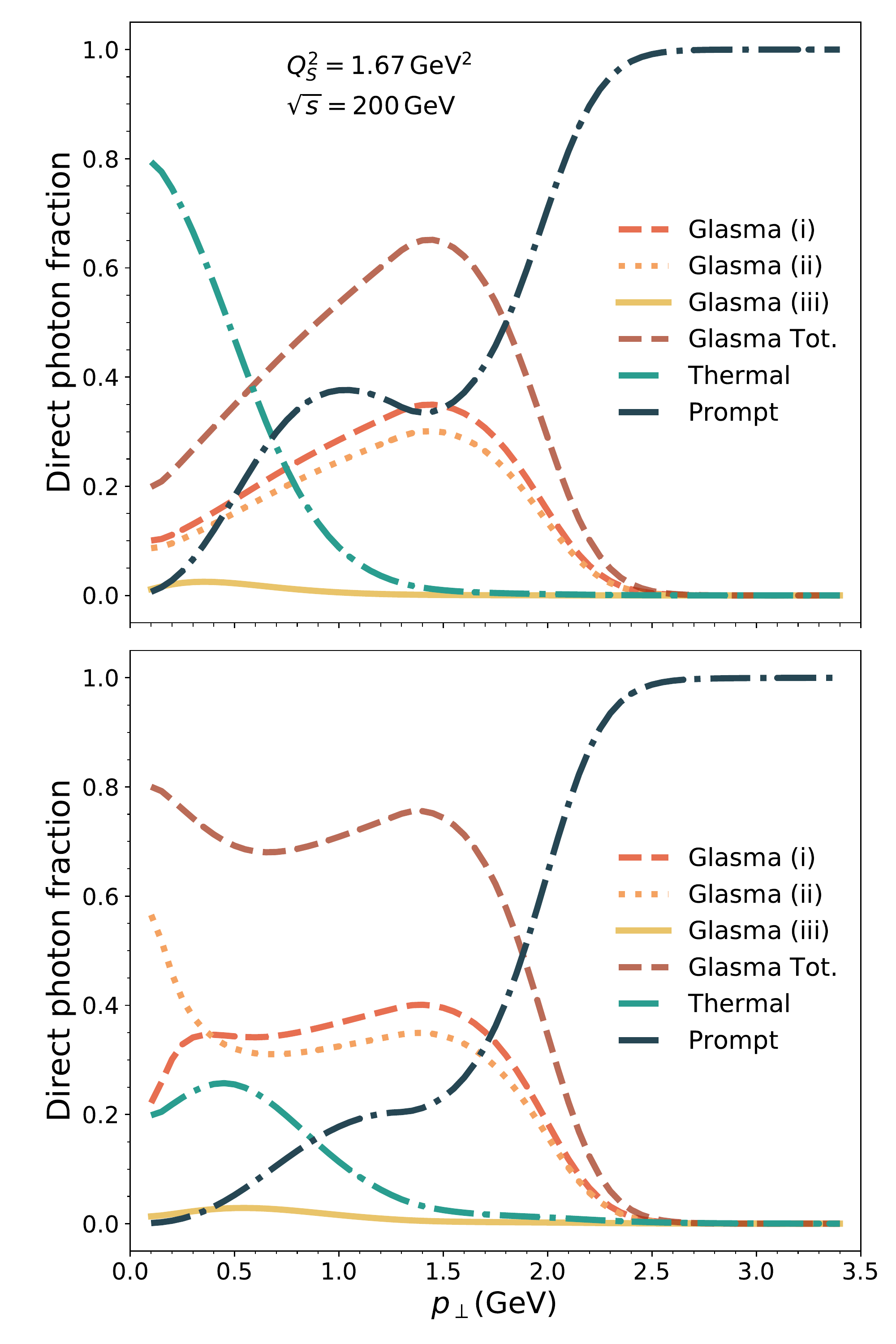}
  \caption{Fraction for each contribution to the total direct photon multiplicity at RHIC energy, $\sqrt{s}=200\,\mathrm{GeV}$, for $0-20\%$ centrality class, and a saturation scale of $Q_s^2=1.67\,\mathrm{GeV}^2$.\textit{Up}: LL fractions for the Glasma and QGP multiplicities.   \textit{Down}: Glasma and QGP multiplicities taken from their LO rates. }
  \label{fig:IRH}
\end{figure}

\begin{figure*}[t]
  \centering
  \includegraphics[width=0.67\textwidth]{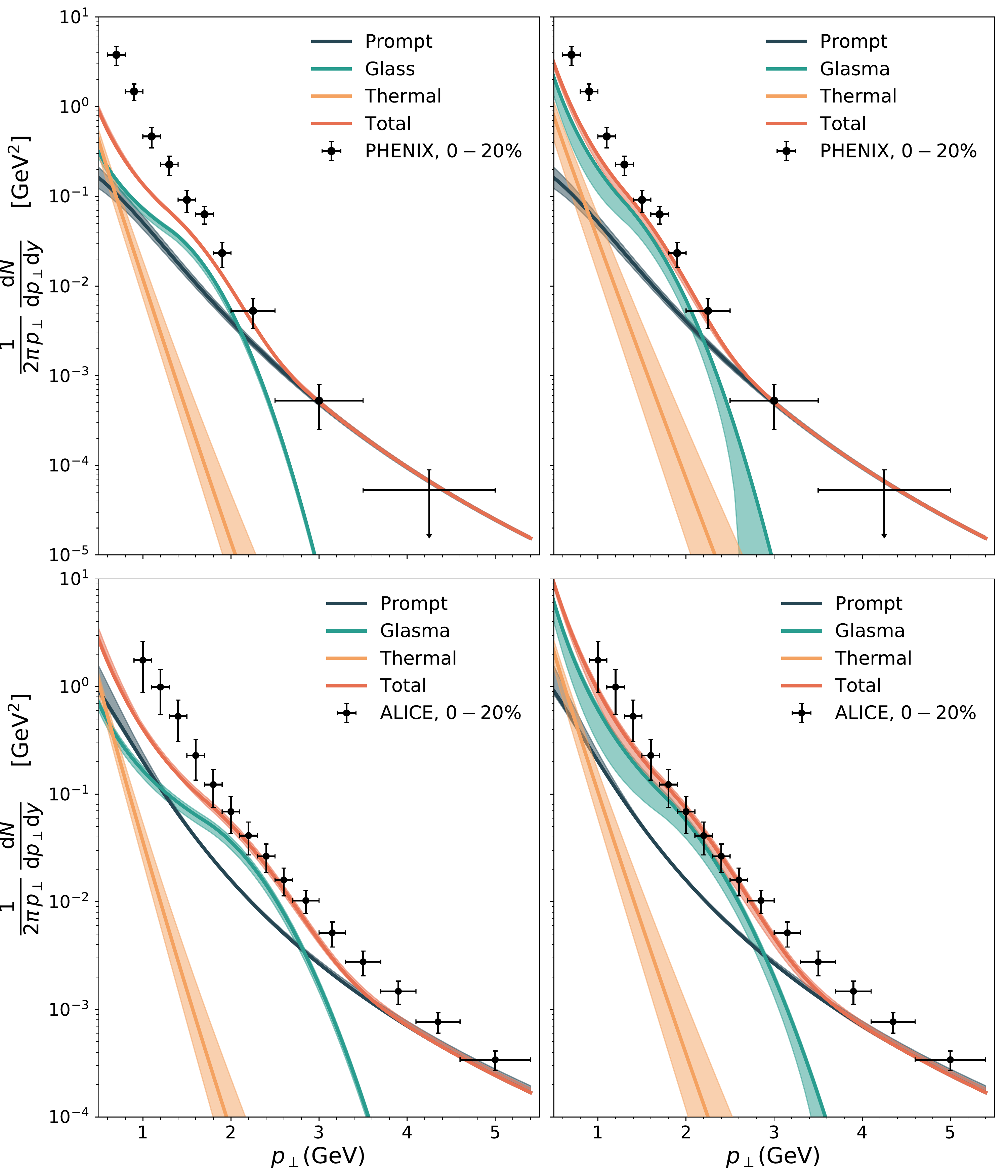}
  \caption{Comparison of the model, BMSS+thermal to experimental data from RHIC \cite{Adare:2014fwh} and LHC \cite{Adam:2015lda}. \textit{Upper left and right}: Comparison of the model with the yield from the $2\leftrightarrow 2$ and collinearly enhanced rates, respectively, at $\sqrt{s}=200\,\mathrm{GeV}$, and $Q_s^2=1.67\,\mathrm{GeV}^2$.  \textit{Lower left and right}: Comparison of the model with the yield from the $2\leftrightarrow 2$ and collinearly enhanced rates, respectively, at $\sqrt{s}=2.76\,\mathrm{TeV}$, and $Q_s^2=2.89\,\mathrm{GeV}^2$.  }
\label{fig:PHENALICE}
\end{figure*}

I have also included the fitting prefactors $K^\gamma$ to account for the total uncertainties of the parameters which may amount for overall normalization discrepancies, such as the initial volume and its evolution, as well as uncertainties in the spatial dependence of initial energy density and subsequent translation to the thermalization temperature and time. It can be seen in eqs. (\ref{eq:GL11PDAP}), (\ref{eq:GL21PDAP}), as well as for the thermal rate, eq. (\ref{eq:thermal1pd}), that these rates could in principle be sensitive to the inclusion of fluctuations of the spatial profile of $Q_s$. This factor is found to be $K^\gamma =2.8$. The result of this fitting can be seen in fig. \ref{fig:PHENALICE}, where I get fair agreement, particularly for ALICE data.

In the previous estimates \cite{Berges:2017eom}, the difference between the thermal and Glasma total yields was found to be of order $\mathcal{O}(1)$
In this work I confirm analytically these results (see secs \ref{sec:NONEQ} and appendix \ref{sec:Anal}) from the fit, regardless of the choice of parameters, once the matching of the constants has been enforced. Nonetheless,  in the $p_\perp$ differential result, a dominant structure was found for Glasma photons.  I want to make an emphasis on the fact that this structure is a signal from the thermalization process. Although its particular shape is model dependent as it strongly depends on the quark distribution, a more refined calculation from kinetic theory would give a reshuffling, or stretching of this yields, giving a more exponential look, while keeping the same order of magnitude around $Q_s$, since the small angle approximation is a good approximation for the hard scatterings of the $2\leftrightarrow 2$ processes in the Glasma. 

For the LL results, this structure is dominant at higher energies, and it is peaked at $p_\perp\sim Q_s$, while the thermal case strongly dominates at small energies, as can be seen in fig. \ref{fig:IRH}. However, once enhanced to the LO rate this is not the case, as Glasma photons completely dominate over the thermal rates. In a realistic simulation, this may become less apparent, as the radial flow from the hydrodynamical expansion will blue-shift the thermal spectrum \cite{Paquet:2017wji}, changing its slope, as well as enhancing the number of photons in this kinematic window.

Apart from a clear-cut comparison between the stages of the evolution of the fireball, the LO version of fig \ref{fig:IRH} allows us also to see the BMSS scenario in action, specifically from the curves for the first and second stages of the Glasma. The direct photon fraction for the first stage exhibits, in the very low-$p_\perp$ limit, a dip compared to its second stage counterpart. This comes directly from the screening mass time-dependence given in eq. (\ref{eq:debye2}). We can see that the rise of soft gluons impacts the system such that it enhances the production of low-$p_\perp$ photons. In a violently evolving Glasma, the overoccupied gluons enhance the number of scatterings, which gives the conditions for photon production. 

As it can be seen in fig. \ref{fig:PHENALICE} and was stated above, in this model the direct photon contribution is dominated by the Glasma, while the thermal contribution is relevant only in the deeper infrared part of the kinematic window $p_\perp\lesssim Q_s$. This happens because of the BMSS estimate of $\tau_{th}$ and $T_{th}$, which gives a late and quite \textit{colder} thermalization, both in RHIC and LHC. One has to remember that in more refined calculations, this may change, since the BMSS scenario poses as a upper bound for the thermalization time. Nonetheless, the calculation here serves as a proof of concept, showing that photons are extremely relevant and may serve a as extra source which may help solve the \textit{photon puzzle}. This has to be contrasted, of course, with current phenomenological ideas, regarding the extra photons to come from later times \cite{vanHees:2014ida}. A comparison as such stresses the need for higher correlation functions, especially those which are more sensitive to the space time evolution. Photon correlations, i.e. Hanbury-Brown-Twiss correlations \cite{Slotta:1996cf,Chapman:1994xa,Gyulassy:1979yi}, may be able to help us discriminate between the two different scenarios\footnote{A manuscript on photon HBT as a way to discern between scenarios is already in preparation \cite{Garcia-Montero:2019kjk}.}.

\begin{figure}[t]
  \centering
  \includegraphics[width=0.47\textwidth]{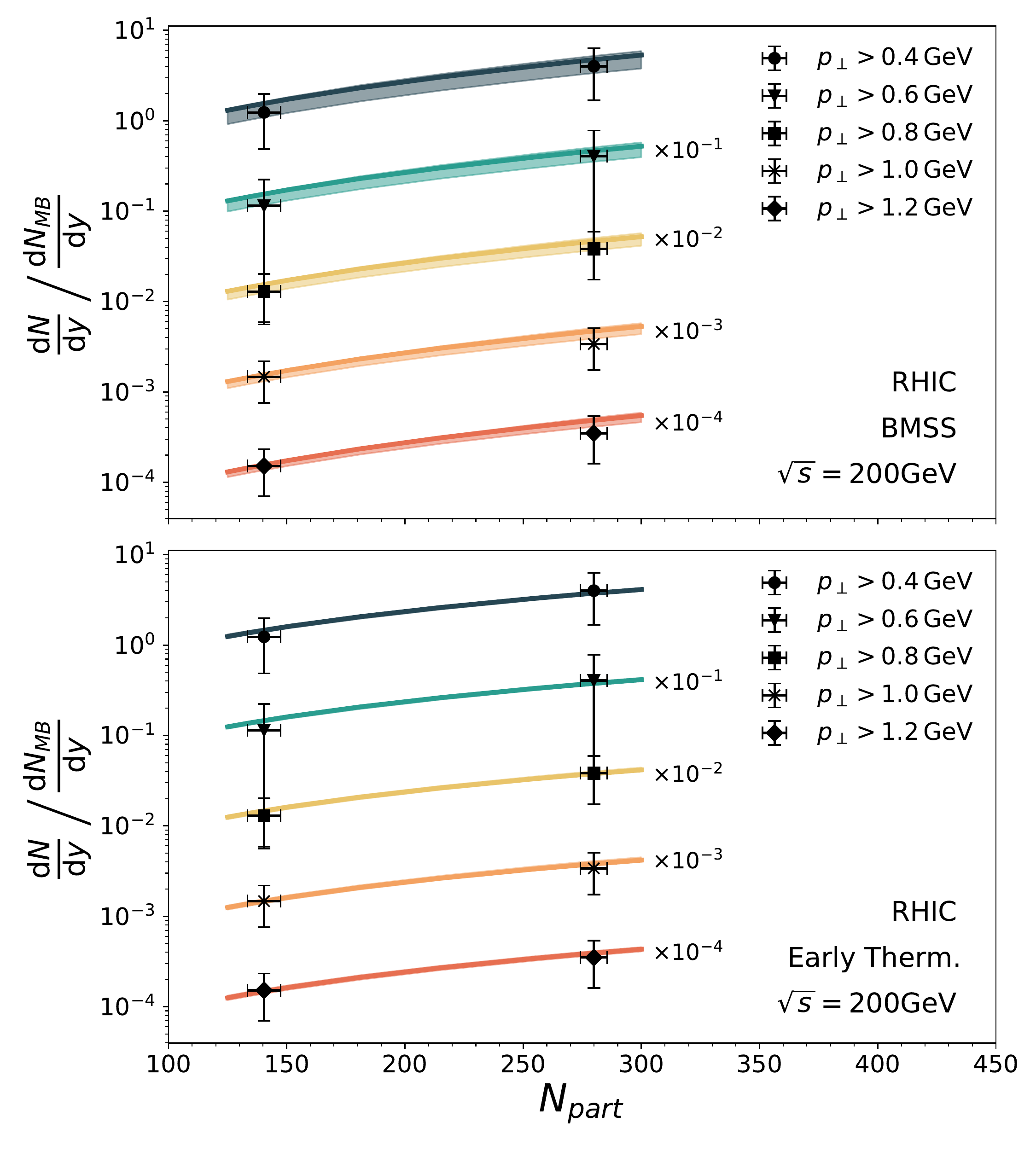}
  \caption{Comparison of the total yield of the Prompt+BMSS+Late Thermal (\textit{up}) and  Prompt+Early+Late Thermal (\textit{down}) scenarios to experimental data in terms of $N_{part}$ \cite{Adare:2014fwh}. Here, the total yield is normalized by its minimum bias value ($N_{part}=106\pm 5.0$). }
\label{fig:glasdndy}
\end{figure}

Aside from the comparisons with the invariant yield from fig.\ref{fig:PHENALICE}, I want to compare also to the total number of photons, in a way which total normalization drops off. For this, I will use the
total yield per unit rapidity, taken as a sum of all momenta, with an infrared cutoff of $p_0$. This quantity in the language of this work is given by  
\begin{equation}
\frac{\rmd  N_{p_0}}{\rmd y}\equiv 2\,\pi\,\int_{p_0}^{\infty} \rmd\,p_\perp\, p_\perp\, \frac{\mathrm{d}N_\gamma }{\mathrm{d}^2 p_\perp\mathrm{d}y}
\label{eq:dndy}
\end{equation}

and will be normalized by its minimum bias counterpart.  For this observable at PHENIX (see fig. \ref{fig:glasdndy}) I observe excellent agreement with the data except for deviations from this trend at $p_0=1.2\mathrm{GeV}$. Furthermore, I checked numerically that for the LO case this observable is completely independent of the choosing $r$, within reasonable ranges given softly by constrain of the data from \cite{Adam:2015lda, Adare:2014fwh}.

\subsection{Early Hydro}

For the phenomenologically matched system the \textit{bottom-up} scenario gives a thermalization time and temperature in the order of $\tau_{th}\sim 2\,\mathrm{fm}$ amd $T_{th}\sim 0.2\, \mathrm{GeV}$. This contrasts with how normally hydrodynamical simulations are initialized, with initial times down to $\tau_i \sim 0.4-0.6\,\mathrm{fm}$ and average initial temperatures as high as $\langle T\rangle_i \sim 0.4-0.6 \mathrm{GeV}$. In this section we will compare the Glasma spectrum with a system in which thermalization occurs at $Q_s\,\tau_0\sim 1$. For this comparison, we will call \textit{Early thermalization} the integrated QGP rate from $\tau_0$ until $\tau_{th}$, for a Bjorken expanding fireball. For the results of this section, the hadronic bremsstrahlung is not taken on account, since for this discussion one is mostly interested in what happens before the system arrives to the freeze-out temperature. 

\begin{figure*}[t]
\centering
  \includegraphics[width=0.95\textwidth]{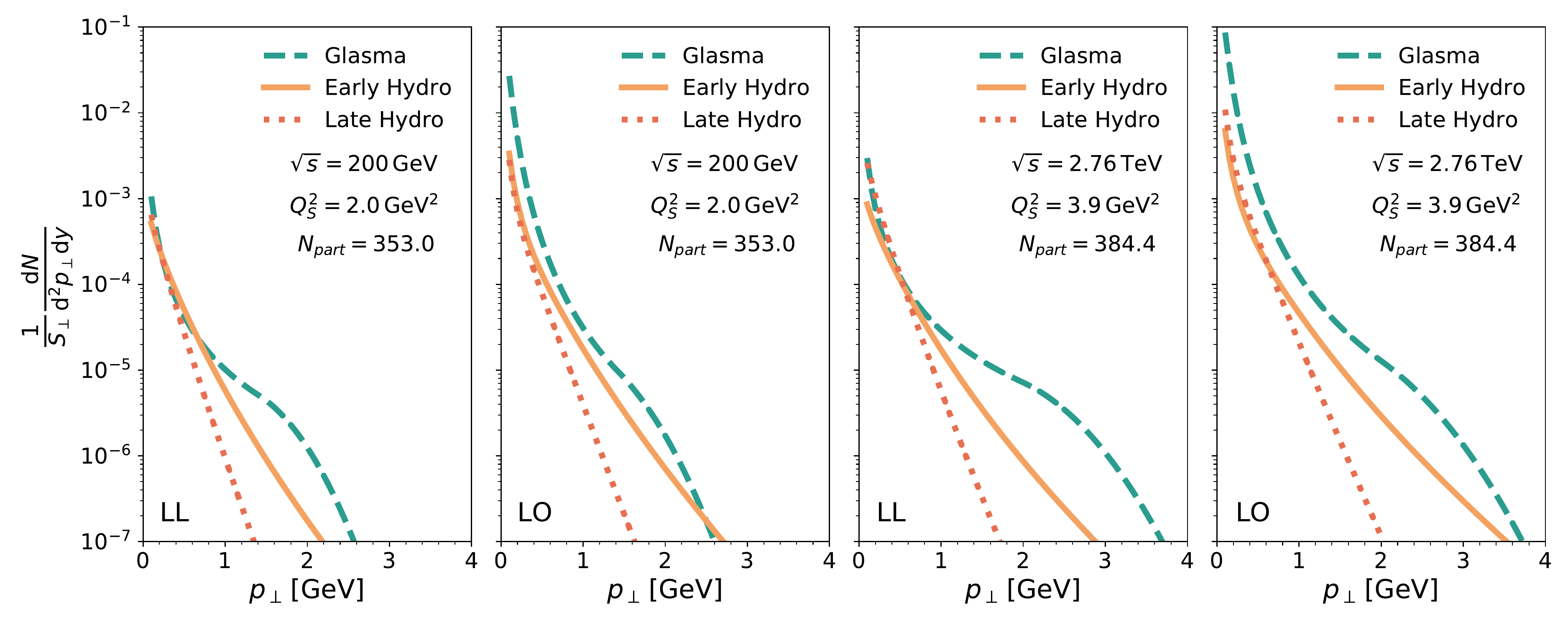}
  \caption{Comparison of BMSS scenario with early thermalization at $0-5\%$ centrality class for RHIC and LHC energies. \textit{Leftmost}: small angle approximation (LL) comparison at $sqrt{s}=200\,\mathrm{GeV}$.\textit{Center left}: LO results at RHIC. The infrared enhancement from section \ref{sec:IRE}  makes the BMSS scenario dominant. \textit{Center right}: LL invariant yields at $sqrt{s}=2.76\,\mathrm{TeV}$. \textit{Right}: LO results at RHIC. The infrared enhancement from section \ref{sec:IRE}  makes the BMSS scenario completely dominant in the kinetic window of interest. }
  \label{fig:EH}
\end{figure*}

In fig. \ref{fig:EH} the reader can see a comparison of the early thermalization scenario with its Glasma counterpart. As a reference, I also give a curve for the late contribution from the thermal QGP. It is immediately apparent that in all results, the structure at $p_\perp\sim Q_s$ dominates, and it is higher than the early hydro contribution. I do not expect the slope of the latter contribution to change significantly with a full-fledged simulation, since the initial conditions normally assigned require the pressure gradients to build up to start the expansion, which will result in the eventual blueshift of the spectrum.

In the case of the LL results, the three contributions deliver similar results for the total photon yield, eq. (\ref{eq:dndy}), as they have similar dependences at the infrared level. This is, however, not the case for the collinearly enhanced result (LO) in which, even though the early hydro and Glasma contributions are still of the same order, the Glasma dominates. This case is particularly strong for higher energies, in which the results of this paper seem to suggest that thermalizing matter is more relevant for the production of photons. In a higher energy system like Pb-Pb at $\sqrt{s}=2,76\,\mathrm{TeV}$ in ALICE, gluons are generally speaking, harder. Apart from increasing the overall order of the radiation, this means that all the system will be shifted towards higher energies, which allows for the near-collinear radiation to be greatly enhanced for the Glasma. 

 On a final note, it is commonly argued \cite{Paquet:2015lta,Paquet:2017wji} that the hydro expansion has to be initialized early on for the pressure gradients to build up, which will cause the blueshift expected from radial flow, as well to account for the transverse anisotropy coefficients, $v_n$. However, this assumption lays on top of a specific pick of initial conditions for the evolution, starting from an initial energy density, without any pre-equilibrium evolution. For a better understanding, a study of the fluctuations of the non-thermal evolution transverse pressure has to be done to start shedding light into this matter. Furthermore, a comprehensive matching of the resulting effective kinetic theory would give the correct initial conditions for hydro evolution, which, conceivably, would start from non-zero transverse velocities. Thus, the assumption that a late thermalization scenario cannot account for the anisotropy coefficients has to be checked via real-time lattice simulations, as it may not be correct. 

\section{Conclusions}

In this work, I derived the photon invariant yield from the Glasma using the  \textit{bottom-up} thermalization framework. My results expand on the parametric estimates for the total photon yields from ref. \cite{Berges:2017eom} by obtaining analytically a transverse momentum resolution for the leading log spectra. In this scenario, a sizeable contribution coming from early times was found, which dominates the total direct photon spectra for $p_\perp\sim Q_s$ and realistic parameters at LHC and RHIC energies. The LL results fail, however, to properly describe the low $p_\perp$ window of the measured spectra. The aforementioned results where improved by using the complete LO thermal rates, as well as introducing a bremsstrahlung modification ansatz for the non-equilibrium rates. 
With this ansatz, the non-thermal contribution becomes mostly dominant and a fair comparison to data can be found up to the overall normalization. To account for this uncertainty an order $\mathcal{O}(1)$ parameter is introduced. 

Apart from the overall normalization, this model is very sensitive to the specific shape of the quark distribution. This uncertainty is just an extension of weak coupling methods to assess collective QCD at RHIC and LHC energies. Nonetheless, this simple model serves as a proof of concept to challenge the idea that pre-equilibrium photon production is suppressed. Nonetheless, as was stated in the text, it was found that the total number of photons is affected only by the assumption of scaling in the quark distribution and not by the fit parameters $\sigma_0$ and $r$, and thus both remain free -yet physically motivated- parameters. This issue may be solved by more thorough calculations, in the context of classical statistical simulations or kinetic theory \cite{Mazeliauskas:2018yef, Tanji:2017xiw,Kurkela:2018wud}. Even if weak coupling assumption has to be kept, this model can be improved by directly fitting  the quark distributions from such more involved calculations. 

As it was stated before, the inclusion of anisotropies is beyond the scope of this work, as I have taken all the distributions to be isotropic and homogeneous in the transverse plane. This approximation, while useful for building this model, relies only on the assumption that flow is suppressed at early times, and its accuracy is not clear in the extreme non-equilibrium setting of the early stages of a HIC. The inclusion of such anisotropies can in principle be added to a simplified model like this one, and lies within the scope of future investigations. 

Even though the BMSS scenario poses as an estimate for the upper limit of the thermalization time, I  would like to argue that this paper presents theoretical evidence that pre-equilibrium photons may be  essential to fully understand the photon spectra recovered from heavy ion collisions. These results, as well as other previous works \cite{Khachatryan:2018ori,McLerran:2014hza,Monnai:2019vup}, seem to suggest that the  \textit{direct photon puzzle} may be solvable by the thorough and 
spatially resolved calculation of non-thermally sourced photons.

\section*{Acknowledgements}
I thank J\"urgen Berges, Eduardo Grossi, Aleksas Mazeliauskas, Jorge Lopez, David Lafferty, Nicole Martin, Klaus Reygers, Naoto Tanji for fruitful discussions. I would like to thank Prithwish Tribedy for giving me the IP-Glasma saturation scale data, as well as Werner Vogelsang for sharing his pQCD calculation. 
This work is part of and supported by the DFG Collaborative Research Centre "SFB 1225 (ISOQUANT)". 
\appendix
\section{Analytical derivation of $2\leftrightarrow 2$ yields}
\label{sec:Anal}
This Appendix works as a brief guide to the calculation of the LL Glasma yields. I will only include here the derivation of stages (i) and (ii) of the BMSS scenario, since stage (iii) was briefly explained in sec. \ref{sec:NONEQ}.\subsection{Stage I}
For the calculation of the invariant yield of the first stage one starts with the space-time integration of the rate in eq. (\ref{noneq1}),
\begin{align}
\frac{1}{S_\perp}\frac{\mathrm{d}N_\gamma }{\mathrm{d}^2 p_\perp\mathrm{d}y} =&\GAMMA\,Q_s^2\, \int^{\tau_1}_{\tau_0}\mathrm{d}\tau\,\tau  \int^{\infty}_{-\infty}\,
\mathrm{d}\eta\\
&\quad\times \frac{1}{Q_s\,\tau}\,f_q(\tau,p_\perp, p_z).\nonumber
\label{eq:RATE1}
\end{align}
Here, the transverse dependence is taken to be homogenous, and parametrised by $S_\perp$, which can be taken then from the Glauber model.  By taking $p_z/p_\perp= \, \sinh(y-\eta) \equiv v$, and defining the dimensionless time $\tilde{\tau} \equiv Q_s\,\tau$ one can find 
\begin{align}
\frac{1}{S_\perp}\frac{\mathrm{d}N_\gamma }{\mathrm{d}^2 p_\perp\mathrm{d}y} =&\GAMMA\, f_0\,\frac{Q_s}{p_\perp}\,\mathcal{I}_{2/3}(p_\perp) \WR\,,
\end{align}
where the unitless function $\mathcal{I}_{a}(p_\perp)$ is given by 
\begin{equation}
\mathcal{I}_a(p_\perp) =\int^{\tilde{\tau}_1}_{\tilde{\tau}_0}\frac{\mathrm{d}\tilde{\tau}}{\tilde{\tau}^{-a} } \int^{\infty}_{-\infty}\,
\frac{\mathrm{d} v}{\sqrt{1+v^2 }} e^{-\frac{1}{2}\left(\frac{p_\perp v}{\sigma(\tilde{\tau})}\right)^2}],.
\end{equation}
The $v$ integral can be performed to find the expression
\begin{equation}
  \int^{\infty}_{-\infty}\,
\frac{\mathrm{d} v}{\sqrt{1+v^2 }} e^{-\frac{1}{2}\left(\frac{p_\perp v}{\sigma(\tilde{\tau})}\right)^2} =e^{\frac{1}{4}\left(\frac{p_\perp }{\sigma(\tilde{\tau})}\right)^2} K_0\left(\frac{p^2}{4\,\sigma^2(\tilde{\tau})}\right)\,,
\end{equation}
where the $\mathcal{I}$ function can be fully integrated to find the invariant yield,
\begin{equation}
\begin{aligned}
\frac{1}{S_\perp}\frac{\mathrm{d}N_\gamma }{\mathrm{d}^2 p_\perp\mathrm{d}y} =&  \frac{10}{3\pi^4} \frac{\alpha_e\,\kappa_g}{\sqrt{2\pi}} \,f_0\, \frac{\sigma_0}{Q_s} \, \WR  \, \frac{Q_s^2}{p^2_\perp}\, \\ 
 &\quad \times \left.  G ^{2,2}_{2,3} \left( \frac{p^2_\perp }{2\,\sigma^2(\tau)}\left|\begin{array}{c} 1\,,\, 1 \\ 1/2,\,1/2,\,0 \end{array}\right. \right) \right|_{\tau_0}^{\tau_1}\,.
\end{aligned}
 \label{eq:GS2MG}
\end{equation}

Here, $G$ stands for the Meijer-G function (see Ref. \cite{gradshteyn2007}). By keeping $p_\perp$ fixed and expanding up to leading order in $\sigma_0/p_\perp$ and substituting $Q_s\,\tau_0=1 $ and  $Q_s\,\tau_1=\alpha_s^{-3/2}$, one can obtain the simplified and more meaningful expression for the photon $p_\perp$ resolved multiplicity. This gives equation (\ref{eq:GL11PDAP}), namely

\begin{equation}
\begin{aligned}
\frac{1}{S_\perp}\frac{\mathrm{d}N_\gamma }{\mathrm{d}^2 p_\perp\mathrm{d}y} =  \sqrt{\frac{\pi}{2}}\frac{20}{9\pi^4} \alpha_e\,\kappa_g & \,f_0\, \frac{\sigma_0}{Q_s} \,\log\left(\alpha_s^{-3/2}\right) \\ 
&\times\, \frac{Q^2_s}{p^2_\perp}\, W_r[p_\perp-Q_s] \,,
\end{aligned}\,.
\end{equation}

As it  was stated in the main text, this expansion can be trusted for values corresponding to $p_\perp > \sigma_0$. Below that, the approximation fails and, in fact, gives a diverging total yield. To find the  yield per unit rapidity, $\mathrm{d}N_\gamma /\mathrm{d}y$, the full result, eq. (\ref{eq:GS2MG}), has to be integrated. This gives the function 

\begin{equation}
\begin{aligned}
\left.\frac{1}{S_\perp}\frac{\mathrm{d}N_\gamma }{\mathrm{d}y}\right|_{p_\perp<Q_s} =&  \frac{10}{3\pi^3} \frac{\alpha_e\,\kappa_g}{ \sqrt{2 \pi }}\,\,f_0\, \sigma_0\, Q_s \\
&\times\left.G_{3,4}^{2,3}\left(\left. \frac{p^2_\perp }{2\,\sigma^2(\tau_1)}\right|
\begin{array}{c}
 1,1,1 \\
 \frac{1}{2},\frac{1}{2},0,0 \\
\end{array}
\right)\right|_{\tau_0}^{\tau_1}
\end{aligned}\,.
\end{equation}

Once again, we can expand this expression to find (\ref{eq:TotalYieldI}) at LO in terms of the normalized anisotropy parameter, $\sigma_0/Q_s$, which once again, we state, confirms the results of the former parametric estimate \cite{Berges:2017eom}.
\subsection{Glasma II}
For the second stage of the Glasma, the integration proceeds in the same fashion. We start now with the rate at stage II,  

\begin{align}
\frac{1}{S_\perp}\frac{\mathrm{d}N_\gamma }{\mathrm{d}^2 p_\perp\mathrm{d}y} =&  \GAMMA\,Q_s^2\, \int^{\tau_2}_{\tau_1}\mathrm{d}\tau\,\tau  \int^{\infty}_{-\infty}\,
\mathrm{d}\eta\\
&\quad\times \frac{1}{(Q_s\,\tau)^{1/2}}\,f_q(\tau,p_\perp, p_z)\,.\nonumber
\end{align}
By making the same substitutions as before, one can find the general expression 
\begin{align}
\frac{1}{S_\perp}\frac{\mathrm{d}N_\gamma }{\mathrm{d}^2 p_\perp\mathrm{d}y} =&\GAMMA\, f_0\,\frac{Q_s}{p_\perp}\,\mathcal{I}_{1/2}(p_\perp) \WR\,,
\end{align}
where integrating $\mathcal{I}_{1/2}(p_\perp)$ as above, one can find the total result for the invariant yield of the second stage of BMSS scenario,
\begin{equation}
\begin{aligned}
\frac{1}{S_\perp}\frac{\mathrm{d}N_\gamma }{\mathrm{d}^2 p_\perp\mathrm{d}y} =&  \frac{10}{3\pi^4} \frac{\alpha_e\,\kappa_g}{2^{5/4}  \sqrt{\pi }}\,  \,f_0 \, \frac{\sigma_0}{Q_s} \, W_r[p_\perp-Q_s] \, \frac{Q_s^2}{p^2_\perp}\, \\ 
 & \quad\times \left.  G_{2,3}^{2,2}\left(\left.\frac{ p_\perp^2}{2 \sigma^2(\tau)}\right|
\begin{array}{c}
 1,\frac{7}{4} \\
 \frac{5}{4},\frac{5}{4},0 \\
\end{array}
\right) \right|_{\tau_1}^{\tau_2}\,.
\end{aligned}
\label{eq:fullG2}
\end{equation}

To find a simplified version of the rate for the kinematic window of interest, we expand in terms of $\sigma_0/p_\perp$ and substitute $\tilde{\tau}_1 = \alpha_s^{3/2}$ and $\tilde{\tau}_2= \alpha_s^{5/2}$  find eq. (\ref{eq:GL21PDAP}), namely
\begin{equation}
\begin{aligned}
\frac{1}{S_\perp}\frac{\mathrm{d}N_\gamma }{\mathrm{d}^2 p_\perp\mathrm{d}y} = & \sqrt{2\,\pi}\frac{20}{9\pi^4} \alpha_e\,\kappa_g\,f_0\, \frac{\sigma_0}{Q_s}  \, \frac{Q^2_s}{p^2_\perp}\, \\ 
	& \times W_r[p_\perp-Q_s] \,\left(\alpha_s^{-1/2}-1\right)\,.
\end{aligned}
\end{equation}

Just as with stage (i), this yield is divergent at low-$p_\perp$ and to be able to find the total yield, one has to integrate eq. (\ref{eq:fullG2}). I find the expression

\begin{align}
\left.\frac{1}{S_\perp\,Q^{2}_S}\frac{\mathrm{d}N_\gamma }{\mathrm{d}y}\right|_{p_\perp<Q_s} =&   \frac{20}{3\pi^3} \frac{\alpha_e\,\kappa_g}{2^{3/4}  \sqrt{\pi }}\,f_0\, \left(\frac{\sigma_0}{ Q_s}\right)^{5/2} \\
&\times\left.G_{3,4}^{2,3}\left(\left. \frac{1 }{2\,\sigma^2(\tau)}\right|
\begin{array}{c}
 1,\frac{7}{4},\frac{7}{4} \\
 \frac{5}{2},\frac{5}{2},0,\frac{3}{4} \\
\end{array}
\right)\right|_{\tau_1}^{\tau_2}\nonumber.
\end{align}

From which, once again, I can confirm the parametrical results from \cite{Berges:2017eom} by expanding in terms of  $\sigma_0/Q_s$, as in eq. (\ref{eq:S2TY})
\section{Thermal Photons}
\label{sec:TH}

For photon production from a thermal medium, I may further simplify the computations 
by using thermal Boltzmann distributions for quarks and gluons at high energies in (\ref{eq:smallangle}). This leads to 
\begin{equation}
E\,\frac{\mathrm{d}N^{\mathrm{th}}}{\mathrm{d}^4X \mathrm{d}^3p} = C \frac{5}{9}\frac{\aem\as}{2\pi^2}T^2 e^{-E/T} \, .
\label{eq:thermalsmallangle}
\end{equation}

From this expression the photon multiplicity is obtained by integrating (\ref{eq:thermalsmallangle}) over the four-volume of the evolution,
\begin{equation}
\frac{\mathrm{d}N^{\mathrm{th}}}{\mathrm{d}y\mathrm{d}^2 p_\perp}  = C \frac{5}{9}\frac{\aem\as}{2\pi^2}\int \rmd\tau\,\rmd\eta\,\rmd^2 x_\perp \, \tau\, T^2\, e^{-p_\perp \cosh(\eta-y)/T} \,.
\end{equation}
Here, I made the equation explicitly Lorentz invariant by 
$E \rightarrow p^\mu u_\mu = p_\perp \cosh(\eta-y)$ 
with the comoving four-velocity $u_\mu = (\cosh \eta, 0,0 \sinh \eta)$. 

\begin{figure}[t!]
\centering
  \includegraphics[width=0.45\textwidth]{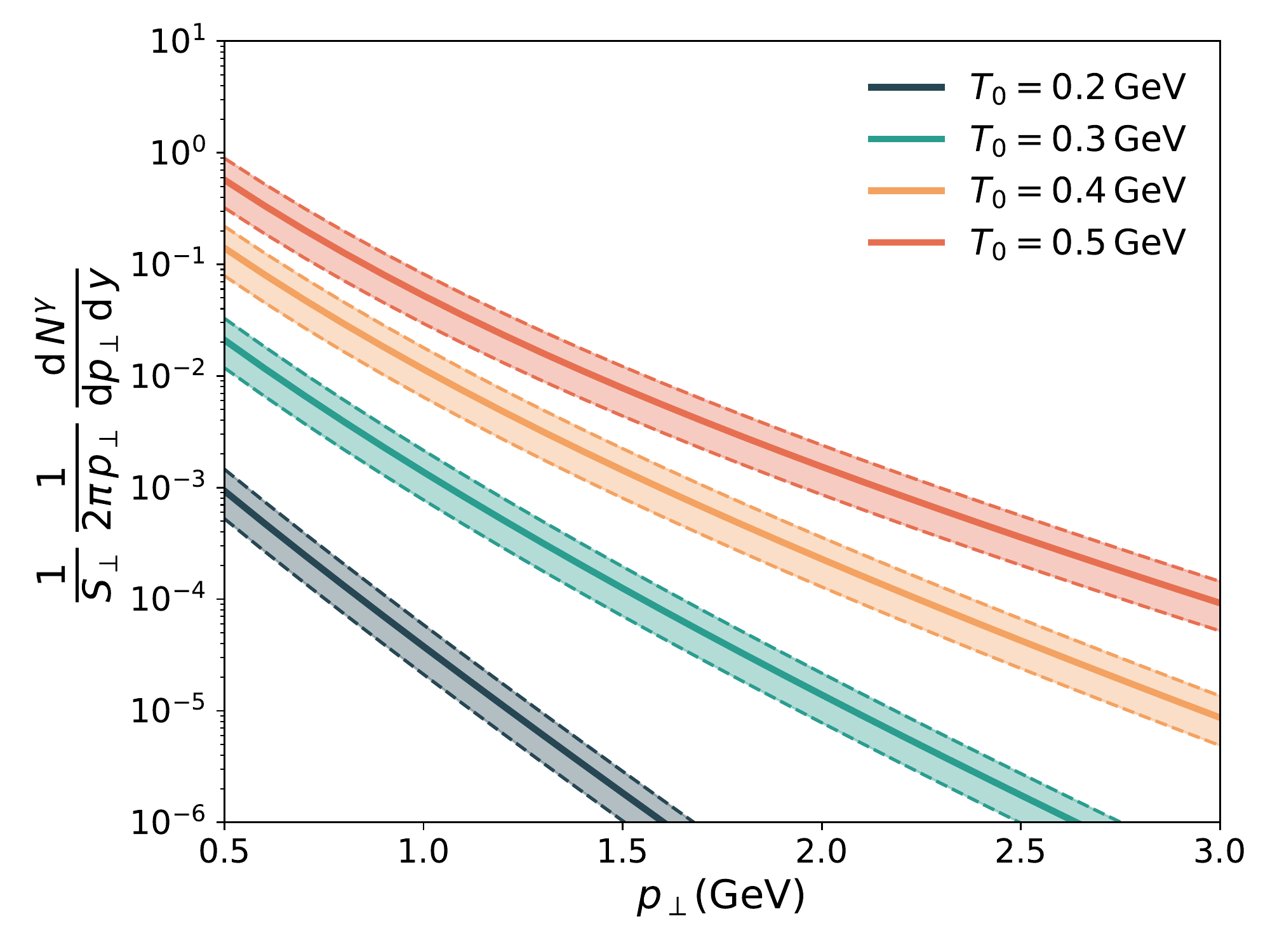}
  \caption{Thermal multiplicities for different thermalization temperature, $T_{th}$,  in terms of $\PTS{}$. In all the curves, the thermalization time has been fixed to $\tau_{th}=2\,\mathrm{fm}$, while the error bands correspond to $50\%$ variation over the thermalization time $\tau_{th}$.}
  \label{fig:NTH}
\end{figure}

The integration itself depends on the spacetime temperature profile. This can be derived from simplified hydrodynamical models or from direct simulation. The simplest scenario would be the evolution for a system invariant under $\eta$, $\xt$ and $\phi$ transformations. From basic symmetry arguments, one can get the evolution of the energy density, 
\begin{equation}
\epsilon = \epsilon_\mathrm{th} \left[\frac{\tau_\mathrm{th}}{\tau}\right]^{4/3}\,.
\end{equation}
Using that $\epsilon \sim T^4$, one gets the $\tau$-dependent temperature profile of the expansion, 
\begin{equation}
T(\tau) = T_\mathrm{th} \left[\frac{\tau_\mathrm{th}}{\tau}\right]^{1/3} \, ,
\label{eq:T}
\end{equation}
where $T_{th}$ is the thermalization temperature, and $\tau_{th}$ is the proper time at which it thermalizes. In the bottom-up thermalization scenario~\cite{Baier:2000sb}, their parameteric dependence is given by
\begin{align}
T_\mathrm{th} \sim c_T c_\mathrm{eq}\, \alpha_s^{2/5}\, Q_s \,,\\
\tau_\mathrm{th} \sim c_\mathrm{eq}\, \alpha_s^{-13/5}\, Q_s^{-1}\,.\nonumber
\label{eq:def_TTH}
\end{align}
Here, $c_\mathrm{eq}$ is a coefficient of order unity, which arises from the uncertainty from the parametric dependence of the thermalization time, $\tau_\mathrm{th}$. The other coefficient, $c_T$,  is a constant needed to finish to constraint the thermalisation temperature in the BMSS scheme \cite{Baier:2000sb,Baier:2002qv}. These coefficients are constrained in sec. \ref{sec:parfix} using the method in ref. \cite{Berges:2017eom}.
For the thermal epoch, the system will evolve from the thermalization time, until the critical time, $\tau_c$, which signals the arrival to the critical temperature. At this point, the deconfined quark-gluon-plasma phase transitions via a crossover to the hadronic phase. For this work I will take $T_c	= 0.154 \, \mathrm{GeV} $ \cite{Borsanyi:2010cj, Bazavov:2011nk}, while $\tau_c$ can be from the temperature profile as follows

\begin{equation}
\tau_c = \tau_{th}\left(\frac{T_{th}}{T_c}\right)\,.
\end{equation}

To start, I want to derive the photon multiplicity for the thermal case. For this, as it was noted before, I integrate emission function, eq.\ref{eq:thermalsmallangle} ,  over the 4-volume. This gives

\begin{align}
\frac{\mathrm{d}N_\gamma }{\mathrm{d}^2 p_\perp\mathrm{d}y} = S_\perp \, \tilde{C}\,\frac{5}{9}\,\frac{\aem\,\as}{2\pi^2}\int_{\tau_{th}}^{\tau_c} \rmd& \tau \,\tau \int_{-\infty}^{\infty}\rmd\eta\, \tau\, T^2(\tau) \\ 
&\,\times e^{-\PTS{} \cosh[\eta-y]/T(\tau)} \,.\nonumber 
\end{align}

Here, the temperature profile is assumed to be homogeneous in the transverse plane, which yields the the $\boldsymbol{x}_\perp$ integration trivial.  The integration can be continued by the rapidity dependent part of the integrand. Using 
\begin{equation}
2\, K_0(z) = \int_{-\infty}^{\infty}\rmd\eta \, e^{ -z \cosh\eta}\,,
\end{equation}
where $K_n(x)$ stands for the modified Bessel function of order $n$. After this integration, I find 

\begin{equation}
\frac{1}{S_\perp}\frac{\mathrm{d}N_\gamma }{\mathrm{d}^2 p_\perp\mathrm{d}y}  =  \, \frac{5}{9}\,\frac{\aem\,\as}{\pi^2} \int_{\tau_{th}}^{\tau_c} \rmd\tau\,\tau\, T^2(\tau)K_0(\PTS{}/T(\tau))  \,.
\end{equation}

By transforming the integration from proper time to inverse temperature, $\beta=1/T$, one can get the integral 
\begin{align}
\frac{1}{S_\perp}\frac{\mathrm{d}N_\gamma }{\mathrm{d}^2 p_\perp\mathrm{d}y} & =  \, \frac{5}{3}\,\frac{\aem\,\as}{\pi^2} \, \tau^2_{th} \, T^6_{th}\int_{\beta_{th}}^{\beta_c} \rmd\beta\,\beta^3\, K_0(\beta \,\PTS{})\nonumber \\
 \,&\equiv  \,\frac{5}{3}\,\frac{\aem\,\as}{\pi^2} \, \tau^2_{th} \, T^6_{th} \,B_{th}(\PTS{})\,.
\end{align}

The function $B_{th}(\PTS{})$  can be found by analytical integration, to  get 
\small
\begin{align}
B_{th}(\PTS{}) =& \bigg\{ \beta_{th}^4 \bigg[\frac{K_1(\beta_{th}\PTS{})}{\beta_{th}\PTS{}}+2 \frac{K_2(\beta_{th}\PTS{})}{(\beta_{th}\PTS{})^{2}}\bigg] \nonumber\\
&-\beta_{c}^4 \bigg[\frac{K_1(\beta_{c}\PTS{})}{\beta_{c}\PTS{}}+2 \frac{K_2(\beta_{c}\PTS{})}{(\beta_{c}\PTS{})^{2}}\bigg]\,\bigg\},
\label{eq:thermal1pd}
\end{align}
\normalsize
 It is important to note that this function is reliable when $\PTS{}>T$, since the emission function I used had the assumption of $E>>T$. Finally, one can find the total yield by integrating over the transverse momentum 
\begin{align}
\frac{1}{S_\perp}\frac{\mathrm{d}N_\gamma }{\mathrm{d}y} & =  \, \frac{10}{3}\,\frac{\aem\,\as}{\pi} \, \tau^2_{th} \, T^6_{th}\int_{0}^{\infty} \mathrm{d} p_\perp\, p_\perp B_{th}(\PTS{}))\nonumber \\
 \,&\equiv  \,\frac{5}{3}\,\frac{\aem\,\as}{\pi} \, \tau^2_{th} \, T^4_{th} \,\left( \frac{T^{2}_{th}}{T_c^2} -1 \right)\,.
\end{align}
where I find the same yield than in ref. \cite{Berges:2017eom} 
Here, as in the previous estimates, I have integrated from $p_\perp =0 $ instead of $p_\perp = T $, where the formula is valid. Here I quantify the  error to be of order $K_2(1)$, which will give a relative error of $\mathcal{O}(1)$. 

\bibliographystyle{apsrev4-1}
\bibliography{References}

\end{document}